\documentclass[aps,prb]{revtex4}

\usepackage{amsmath,amssymb,mathrsfs}
\usepackage[dvips]{graphicx}
\usepackage{psfrag}

\newcommand{\bs}[1]{\boldsymbol{#1}}
\newcommand{\ket}[1]{\left|#1\right\rangle}
\newcommand{\bra}[1]{\left\langle#1\right|}
\newcommand{\sgn}[1]{\text{sgn}(#1)\,}
\newcommand{\ii}{\text{i}\,}

\def\ie{\emph{i.e.,\ }}
\def\eg{\emph{e.g.\ }}

\allowdisplaybreaks[4]

\begin{document}
\title{Dynamical response functions in the quantum Ising chain with a boundary}
\author{Dirk Schuricht and Fabian H. L. Essler}
\date{June 29, 2007}

\affiliation{The Rudolf Peierls Centre for Theoretical Physics, 
University of Oxford,\\ 1 Keble Road, OX1 3NP, Oxford, United Kingdom}

\pagestyle{plain}

\begin{abstract}
We determine dynamical response functions
$\langle{\cal O}^\dagger(t,x_1){\cal O}(0,x_2)\rangle$ 
in the scaling limit of the
quantum Ising chain on the half line in the presence of a boundary
magnetic field. Using a spectral representation in terms of infinite
volume form factors and a boundary state, we derive an expansion for
the correlator that is found to be rapidly convergent as long as
$|\frac{x_1+x_2}{\xi}|\agt 0.2$ where $\xi$ is the correlation length.
At sufficiently late times we observe oscillatory behaviour of the
correlations arbitrarily far away from the boundary. We investigate the
effects of the boundary bound state that is present for a range of
boundary magnetic fields.
\end{abstract}

\maketitle

\section{Introduction}
Over the last decade there has been significant progress in the calculation of
dynamical correlation functions in integrable quantum field
theories~\cite{Smirnov92book,fring93,Lukyanov95,lukyanov,babujian,Delfino04}.
This has made it possible to determine the dynamic response of a variety of
experimentally relevant one-dimensional models of Ising
magnets~\cite{DelfinoMussardo95,CarrTsvelik03},
integer~\cite{applSint,Konik03} and spin-1/2~\cite{applShalf} quantum magnets,
Mott insulators~\cite{applMI}, two-leg ladders~\cite{appl2leg}, carbon
nanotubes~\cite{applnano} and ultra-cold atomic gases \cite{applgas}. Having
established the bulk behaviour of such systems, it is of interest to
investigate the effects of impurities on the dynamic response. The simplest
possible effect of non-magnetic impurities in a spin chain material is to
break the chains into finite segments.  Resulting changes in the
susceptibility have been investigated in both spin-1/2~\cite{affleck} and
spin-1~\cite{edgestates} Heisenberg chains as well as two-leg Heisenberg
ladders~\cite{ladder}. In the latter two cases the chain break can lead to the
formation of edge states, which have been observed in inelastic neutron
scattering experiments~\cite{edgestates2}.  Impurity and boundary effects are
also relevant to scanning tunneling microscopy experiments~\cite{STM}, where
the associated breaking of translational symmetry can lead to fingerprints in
the dynamic response that assist in characterising of the underlying bulk
behaviour~\cite{Kivelson}. For the aforementioned reasons it is important to
extend the calculation of dynamical response functions in one-dimensional
models to systems with boundaries. In the gapless case this is readily
accomplished by boundary conformal field theory~\cite{Cardy84,Cardy89}, while
the gapful case is as usual more involved. One successful approach has been
the extension of the truncated conformal space approach~\cite{TCSA} to systems
with boundary~\cite{Dorey-98,Dorey-01}. The method we will follow in the
present work is the extension of the form factor bootstrap approach to
boundary integrable field
theories~\cite{GhoshalZamolodchikov94,FringKoeberle94,Konik-96,BajnokGeorge06,
  Bajnok-07}. We will focus on the simplest integrable field theory, the
quantum Ising chain~\cite{Onsager44,McCoyWu73,ising1,Berg-79,Bariev,
  CardyMussardo90,YurovZamolodchikov91,Delfino04}.  One-point functions in the
boundary Ising model have been studied by Konik~\emph{et al.}~\cite{Konik-96}.
The purpose of the present work is to analyse two-point functions relevant to
scattering experiments.

This article is organised as follows. First, we will review the basic facts on
the Ising field theory in the bulk. We proceed by discussing the Ising model
on the half line in the presence of a boundary magnetic field and the form
factor bootstrap approach to boundary integrable field theories. In order to
set the stage, we calculate the Green's functions of the Majorana fermions,
which show typical light-cone effects. After a brief discussion of the local
magnetisation we proceed with our main result, the calculation of the
dynamical spin-spin correlations and the corresponding spectral function. The
derived expansion in terms of infinite volume form factors and a boundary
state is found to be rapidly convergent as long as $|\frac{x_1+x_2}{\xi}|\agt
0.2$ where $\xi$ is the correlation length.  As for the Green's functions we
observe light-cone effects, which result in oscillatory behaviour of the
correlations arbitrarily far away from the boundary. Furthermore, we study the
effects of the boundary bound state, which exists for sufficiently small
values of the boundary magnetic field, on the correlation function. Finally,
we discuss the two-point function of the disorder field, which shows the same
qualitative features as the spin-spin correlations.

\section{Quantum Ising chain}
The Hamiltonian of one-dimensional quantum Ising model is given by
\begin{equation}
\label{eq:Isinglattice}
H_\mathrm{latt}=-J\sum_i \bigl(\sigma^\mathrm{z}_i\,\sigma^\mathrm{z}_{i+1}
+\lambda\,\sigma^\mathrm{x}_i\bigr).
\end{equation}
Here $\sigma^\mathrm{x}$ and $\sigma^\mathrm{z}$ are the Pauli matrices and
$J$ is the exchange energy. The model (\ref{eq:Isinglattice}) is related to
the two-dimensional classical Ising model, see for example
Ref.~\onlinecite{GogolinNersesyanTsvelik98}. The critical temperature of the
latter is related to the coupling constant of the quantum Ising chain by
$\lambda=\frac{T-T_\mathrm{c}}{T_\mathrm{c}}$. The Hamiltonian
\eqref{eq:Isinglattice} is invariant under the $\mathbb{Z}_2$-transformation
$\sigma_i^\mathrm{x}\rightarrow\sigma_i^\mathrm{x}$,
$\sigma_i^\mathrm{z}\rightarrow-\sigma_i^\mathrm{z}$.  For $\lambda<1$, which
corresponds to the low-temperature phase of the 2D classical Ising model, this
symmetry is broken. The order parameter field $\sigma_i^\mathrm{z}$ takes a
non-zero expectation value and the ground state is two-fold degenerate. For
$\lambda>1$ the model has a unique ground state. This regime corresponds to
the disordered high-temperature phase of the classical 2D Ising model. The
point $\lambda=1$ is the location of a quantum phase transition.

At small deviations from criticality, $|\lambda-1|\ll 1$, one can pass to the
continuum
limit~\cite{ItzyksonDrouffe89vol1,GogolinNersesyanTsvelik98}. This
leads to the Ising field theory, which is defined by the Euclidean
action~\cite{ItzyksonDrouffe89vol1,Delfino04} 
\begin{equation}
\mathcal{S}=\frac{1}{2\pi}
\int d^2x \left(\psi\bar{\partial}\psi+\bar{\psi}\partial\bar{\psi}+
\ii M\bar{\psi}\psi\right), 
\label{eq:IFT}
\end{equation}
where $\psi$ and $\bar{\psi}$ are the two components of a Majorana fermion.
The complex coordinates used in \eqref{eq:IFT} are given by $z=\tau+\ii x$ and
$\bar{z}=\tau-\ii x$, where $\tau=\ii t$ denotes imaginary time. This results
in $\partial=\partial_z=(\partial_\tau-\ii \partial_x)/2$ and
$\bar{\partial}=\partial_{\bar{z}}=(\partial_\tau+\ii \partial_x)/2$.  The
mass is proportional to the distance from the critical point, $M\propto
J(1-\lambda)$. The model \eqref{eq:IFT} is conformally invariant at the
critical point $M=0$ (see for example
Ref.~\onlinecite{DiFrancescoMathieuSenechal97}). In the ordered phase, which
we will consider throughout this paper, the mass is positive. Furthermore, we
set the velocity to one and use the short distance normalisations
\begin{equation}
\label{eq:normalizations}
z\bra{0}\psi(\tau,x)\,\psi(0,0)\ket{0}\rightarrow 1,\quad
|z|^{1/4}\bra{0}\sigma^\mathrm{z}(\tau,x)\,
\sigma^\mathrm{z}(0,0)\ket{0}\rightarrow 1,\quad
\text{as}\quad|z|\rightarrow 0.  
\end{equation}

For later use we will need the notion of mutual semi-locality of
operators
\cite{YurovZamolodchikov91,Smirnov92book,Lukyanov95,Delfino04}. Let us
consider the operator product 
$O_1(\tau,x)\,O_2(0,0)=O_1(z,\bar{z})\,O_2(0,0)$. If we take $O_1$
counterclockwise around $O_2$ in the plane, \ie we perform the
analytic continuation $z\rightarrow e^{2\pi\ii}z$, $\bar{z}\rightarrow
e^{-2\pi\ii}\bar{z}$, the operators $O_1$ and $O_2$ are said to be
mutually semi-local if 
\begin{equation}
\label{eq:defsemilocality}
O_1(e^{2\pi\ii}z,e^{-2\pi\ii}\bar{z})\,O_2(0,0)=
l_{O_1O_2}\,O_1(z,\bar{z})\,O_2(0,0).
\end{equation}
The phase $l_{O_1O_2}$ is called the semi-locality factor. The two fields are
mutually local if $l_{O_1O_2}=1$. Semi-locality is the mildest form of
non-locality, in general the right-hand side of \eqref{eq:defsemilocality} may
be more complicated. The mutual semi-locality factor of the spin and
disorder operators can be read off from their operator product
expansion~\cite{DiFrancescoMathieuSenechal97,Delfino04} 
\begin{equation}
\sigma^\mathrm{z}(z,\bar{z})\,\mu^\mathrm{z}(0,0)\sim
\frac{1}{\sqrt{2}\,|z|^{1/4}}\,\bigl(e^{\ii\pi/4}\sqrt{z}\,\psi(0)
+e^{-\ii\pi/4}\sqrt{\bar{z}}\,\bar{\psi}(0)\bigr).
\label{eq:OPE}
\end{equation}
This implies that when taking $\sigma^\mathrm{z}$ once around $\mu^\mathrm{z}$
one obtains an extra minus sign, \ie $l_{\sigma^\mathrm{z}\mu^\mathrm{z}}=-1$.
In the same way one finds
$l_{\psi\mu^\mathrm{z}}=l_{\bar{\psi}\mu^\mathrm{z}}=
l_{\psi\sigma^\mathrm{z}}=l_{\bar{\psi}\sigma^\mathrm{z}}=-1$. On the
other hand, the disorder field $\mu^\mathrm{z}$ is local with respect
to itself. 

In order proceed further, we need to construct a basis of scattering
states. In order to do so, we need to specify which ``fundamental field''
(that is a field that has a non-vanishing matrix element between the
vacuum and the one-particle states) we will use to create the
fundamental excitations. In order to avoid additional signs in many of
the equations it is customary to choose the fundamental field to be bosonic.
We will therefore use the disorder field $\mu^\mathrm{z}$. This
implies that the fundamental excitations are viewed as bosons with
two-particle scattering matrix $S=-1$. Let us denote the corresponding
annihilation and creation operators by $A(\theta)$ and
$A^\dagger(\theta)$ respectively. They fulfil the
Faddeev--Zamolodchikov algebra
\begin{eqnarray}
A(\theta_1)A(\theta_2)&=&SA(\theta_2)A(\theta_1),\nonumber\\*
A^\dagger(\theta_1)A^\dagger(\theta_2)&=&
SA^\dagger(\theta_2)A^\dagger(\theta_1),\label{eq:Aalgebra}\\*
A(\theta_1)A^\dagger(\theta_2)&=&2\pi\delta(\theta_1-\theta_2)
+SA^\dagger(\theta_2)A(\theta_1),\nonumber
\end{eqnarray}
where the scattering matrix is $S=-1$. The vacuum state is then
defined by 
\begin{equation}
A(\theta)|0\rangle=0,
\end{equation}
and a basis of scattering states is given as
\begin{equation}
  \label{eq:defA}
  \ket{\theta_1,\ldots,\theta_n}=
  A^\dagger(\theta_1)\ldots A^\dagger(\theta_n)\ket{0}.
\end{equation}
In terms of the states (\ref{eq:defA}) the resolution of the identity
reads 
\begin{equation}
\mathrm{id}=\ket{0}\bra{0}+\sum_{n=1}^\infty\frac{1}{n!}
\int_{-\infty}^\infty\frac{d\theta_1\ldots d\theta_{n}}{(2\pi)^n}
\ket{\theta_1,\ldots,\theta_n}\bra{\theta_n,\ldots,\theta_1}.
\label{eq:residentity}
\end{equation}
For the calculation of correlation functions the knowledge of the matrix
elements or form factors of local operators is necessary. In the
ordered phase the non-vanishing form factors of $\sigma^\mathrm{z}$
contain an even number of particles and are explicitly given
by~\cite{Berg-79,CardyMussardo90,YurovZamolodchikov91,Delfino04}
\begin{equation}
\label{eq:sigmaff}
f(\theta_1,\ldots,\theta_{2n})=
\bra{0}\sigma^\mathrm{z}(0,0)\ket{\theta_{1},\ldots,\theta_{2n}}=
\ii^n\sigma_0\prod_{\substack{i,j=1\\ i<j}}^{2n}
\tanh\frac{\theta_i-\theta_j}{2},
\end{equation}
where $\sigma_0=\bra{0}\sigma^\mathrm{z}\ket{0}$. For comparison, the
form factors of the disorder operator $\mu^\mathrm{z}$ are only non-vanishing
if the number of particles is odd,
\begin{equation}
\bra{0}\mu^\mathrm{z}(0,0)\ket{\theta_{1},\ldots,\theta_{2n+1}}=
\ii^n\sigma_0\prod_{\substack{i,j=1\\ i<j}}^{2n+1}
\tanh\frac{\theta_i-\theta_j}{2}.
\label{eq:muff} 
\end{equation}
The form factors \eqref{eq:sigmaff} and \eqref{eq:muff} follow from
 the following set of requirements~\cite{Smirnov92book,Lukyanov95,Delfino04}:
\begin{enumerate}
\item The form factors $f(\theta_1,\ldots,\theta_{n})$ are meromorphic
  functions of $\theta_n$ in the physical strip $0\le
  \mathfrak{Im}\,\theta_n\le 2\pi$. There exist only simple poles in this
  strip.
\item Scattering axiom:
\begin{equation*}
f(\theta_1,\ldots,\theta_{i+1},\theta_i,\ldots,\theta_n)=
S\,f(\theta_1,\ldots,\theta_i,\theta_{i+1},\ldots,\theta_n),
\end{equation*}
with the scattering matrix $S=-1$.
\item Periodicity axiom:
\begin{equation*}
f(\theta_1+2\pi\ii,\theta_2,\ldots,\theta_n)=
l_{O \mu^\mathrm{z}}\,f(\theta_2,\ldots,\theta_{n},\theta_1),
\end{equation*}
where $O=\sigma^\mathrm{z}$ or $\mu^\mathrm{z}$. The mutual non-locality
phases are given by $l_{\sigma^\mathrm{z}\mu^\mathrm{z}}=-1$ and
$l_{\mu^\mathrm{z}\mu^\mathrm{z}}=1$, respectively.
\item Lorentz invariance: 
\begin{equation*}
f(\theta_1+\alpha,\ldots,\theta_n+\alpha)=
e^{s\alpha}\,f(\theta_1,\ldots,\theta_n),
\end{equation*}
where $s$ denotes the spin of the fields, which is $s_{\sigma^\mathrm{z}}=0$
and $s_{\mu^\mathrm{z}}=0$ as well as $s_{\psi}=-1/2$ and
$s_{\bar{\psi}}=1/2$.
\item Annihilation pole axiom:
\begin{equation*}
\mathrm{Res}\bigl[f(\theta',\theta,\theta_1,\ldots,\theta_n),
\theta'=\theta+\ii\pi\bigr]=\ii
f(\theta_1,\ldots,\theta_n)
\left[1-l_{O\mu^\mathrm{z}}\prod\nolimits_{i=1}^nS\right].
\end{equation*}
Note that the squared brackets on right-hand side equal $2$ for
$\sigma^\mathrm{z}$ as well as $\mu^\mathrm{z}$, as the extra minus sign due
to $l_{\sigma^\mathrm{z}\mu^\mathrm{z}}=-1$ is compensated by an additional
factor $S$.  As there exist no bound states in the Ising model, these are the
only poles of the form factors.
\end{enumerate}
We note that if we had used the Majorana fermion $\psi$ as fundamental
field, the excitations would be viewed as fermions with unity
scattering matrix. Furthermore, there would appear additional minus
signs in the axioms above~\cite{YurovZamolodchikov91}. 

\section{Boundary Ising model}\label{sec:boundary}
We now turn to the Ising field theory on the half-plane $(\tau,x)$,
$\tau\in\mathbb{R}$, $x\in(-\infty,0]$. The boundary is located at $x=0$ and
$\tau$ denotes imaginary time ($\tau=\ii t$).  The Hilbert space of states
associated with the semi-infinite line $\tau=\text{const.}$, $-\infty<x\le 0$,
is denoted by $\mathcal{H}_\mathrm{B}$. The boundary condition is imposed
through application of an external time-independent magnetic field $h$ which
couples to the boundary spin
$\sigma_\mathrm{b}^\mathrm{z}(\tau)=\sigma^\mathrm{z}(\tau,0)$.
The action for this system is given by~\cite{GhoshalZamolodchikov94}
\begin{equation}
\mathcal{S}
=\frac{1}{2\pi}\int d\tau \int_{-\infty}^0 dx 
\left(\psi\bar{\partial}\psi+\bar{\psi}\partial\bar{\psi}+
\ii M\bar{\psi}\psi\right)
+\frac{1}{2\pi}\int d\tau \left(h\sigma_\mathrm{b}^\mathrm{z}
-\frac{\ii}{2}\bar{\psi}\psi-\frac{1}{2}\,a\,\partial_\tau a\right).
\label{eq:IFTboundary}
\end{equation}
It was shown by Affleck and Ludwig~\cite{AffleckLudwig91} that in the critical
model the application a magnetic field at the boundary generates a flow from
free to fixed boundary conditions. The properties of the boundary spin
operator $\sigma_\mathrm{b}^\mathrm{z}$ were analysed in
Refs~\onlinecite{Cardy89,CardyLewellen91}. In the Lagrangian framework defined
by \eqref{eq:IFTboundary} it can be written as~\cite{GhoshalZamolodchikov94}
\begin{equation}
\label{eq:boundarysigma}
\sigma_\mathrm{b}^\mathrm{z}(\tau)=
\frac{1}{2}\bigl(\psi(\tau,x)+\bar{\psi}(\tau,x)\bigr)\Big|_{x=0}\,a(\tau).
\end{equation}
Here $a(\tau)$ is an additional fermionic boundary degree of freedom which is
introduced to describe the ground-state degeneracy and anticommutes with
$\psi$ and $\bar{\psi}$. The cases $h=0$ and $h\rightarrow\infty$ correspond
to free and fixed boundary conditions, respectively. We note that the
application of a non-zero boundary magnetic field removes the ground-state
degeneracy of the Ising model.

In terms of the Majorana fermions the boundary condition reads
\begin{equation}
\label{eq:boundarymajorana}
-\ii\frac{d}{d\tau}\bigl(\psi-\bar{\psi}\bigr)\Big|_{x=0}=
\frac{h^2}{2}\bigl(\psi+\bar{\psi}\bigr)\Big|_{x=0}.
\end{equation}
As was shown by Ghoshal and Zamolodchikov~\cite{GhoshalZamolodchikov94} the
Ising model with boundary conditions \eqref{eq:boundarymajorana} still
possesses infinitely many integrals of motion (which can be constructed from
the integrals of motion of the model in the bulk) and hence remains
integrable.

\begin{figure}
\psfrag{REP1}{$A^\dagger(\theta_1)$}
\psfrag{REP2}{$A^\dagger(\theta_2)$}
\psfrag{REP3}{$S$}
\psfrag{REP4}{$A^\dagger(\theta)$}
\psfrag{REP4a}{$A^\dagger(-\theta)$}
\psfrag{REP5}{$R(\theta)$}
\includegraphics[scale=0.2]{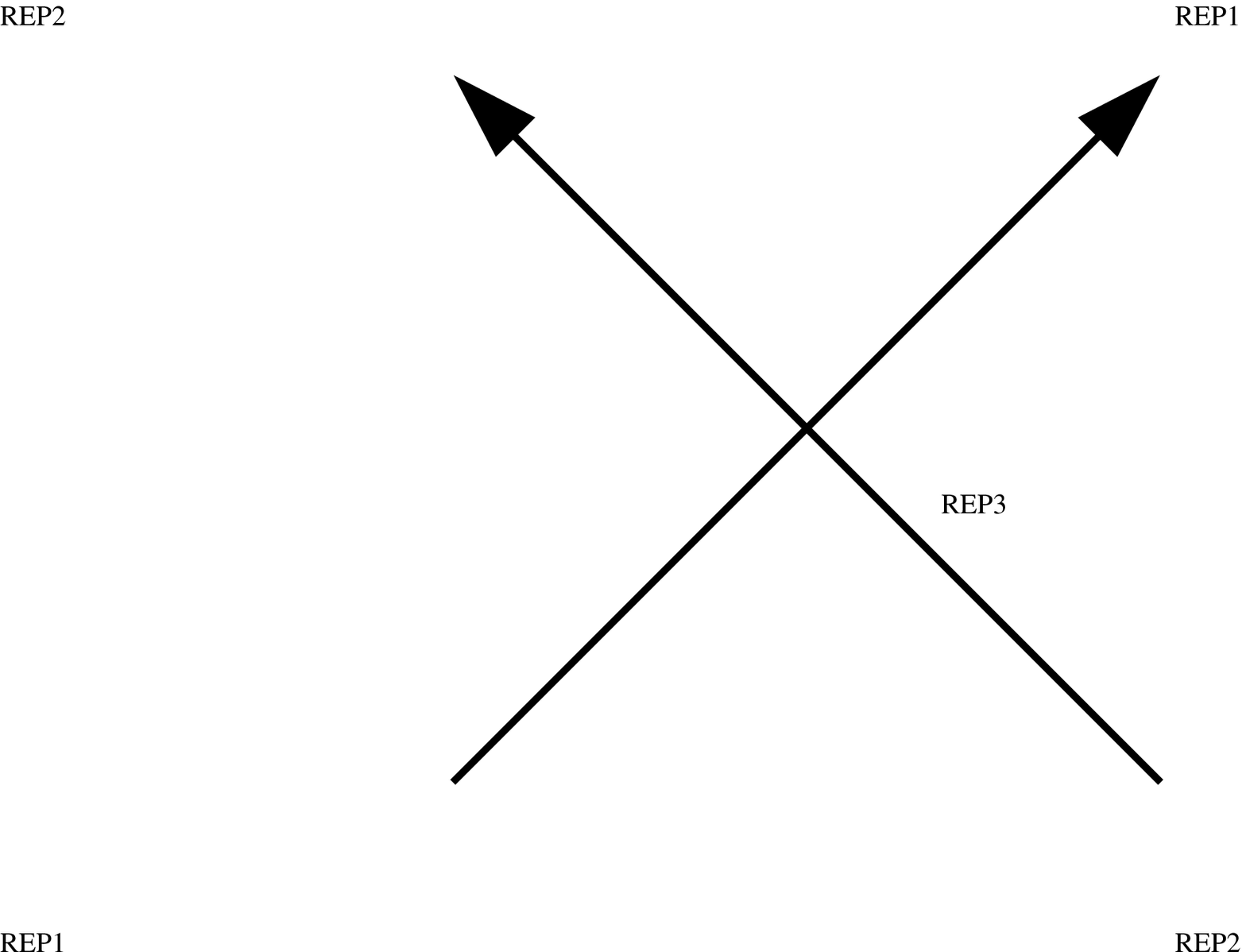}\hspace{30mm}
\includegraphics[scale=0.2]{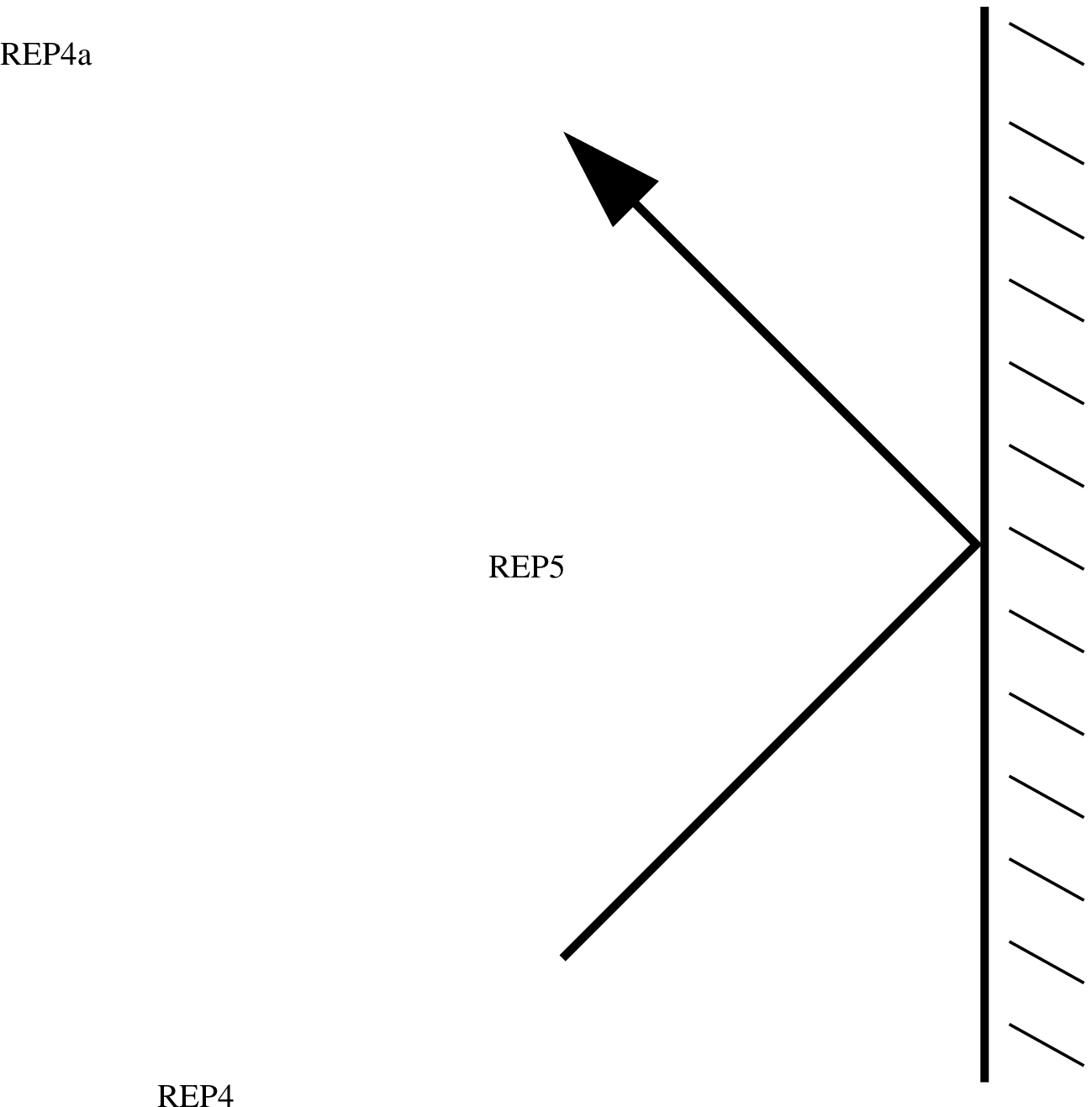}
\caption{Two-particle scattering and scattering off the boundary.}
\end{figure}
If one thinks of the boundary as an infinitely heavy impenetrable particle $B$
sitting at $x=0$ and writes
$\ket{0_\mathrm{B}}=B\ket{0}\in\mathcal{H}_\mathrm{B}$ for the ground state in
the presence of the boundary, the scattering of particles created by the
operators $A^\dagger(\theta)$ at the boundary is encoded in the relations
\begin{equation}
\label{eq:boundaryscattering}
A^\dagger(\theta)B=R(\theta)A^\dagger(-\theta)B.
\end{equation}
The function $R(\theta)$ is interpreted as the single-particle reflection
amplitude off the boundary.  In order to preserve the integrability of the
corresponding bulk system, the boundary scattering matrix $R(\theta)$ has to
satisfy several conditions as discussed in 
Ref.~\onlinecite{GhoshalZamolodchikov94}, which leads to the following
result for the boundary scattering matrix for the Ising model
with boundary magnetic field $h$
\begin{equation}
  \label{eq:Rmatrix}
  R(\theta)=\ii \tanh\!\left(\frac{\ii \pi}{4}-\frac{\theta}{2}\right)
  \frac{\kappa-\ii \sinh\theta}{\kappa+\ii \sinh\theta},\quad
  \kappa=1-\frac{h^2}{2M}.
\end{equation}
Free boundary conditions are recovered for $h=0$, whereas fixed boundary
conditions are obtained in the limit 
$h\rightarrow\infty$ (we restrict ourselves to $h\ge 0$).  
The purpose of the present work is to calculate the two-point
correlation function 
\begin{equation}
\label{eq:cfboundary}
C(\tau,x_1,x_2)=
\bra{0_\mathrm{B}}\mathcal{T}_\tau\,\sigma^\mathrm{z}(\tau,x_1)\,
\sigma^\mathrm{z}(0,x_2)\ket{0_\mathrm{B}},
\end{equation}
where $\mathcal{T}_\tau$ is the time-ordering operator. The time-dependence
of the operators $\sigma^\mathrm{z}$ is given by
\begin{equation}
\label{eq:time1}
\sigma^\mathrm{z}(\tau,x)=
e^{\tau H_\mathrm{B}}\,\sigma^\mathrm{z}(0,x)\,e^{-\tau H_\mathrm{B}},
\end{equation}
where $H_\mathrm{B}$ is the Hamiltonian of the system in the presence of the
boundary.

\begin{figure}
\psfrag{REP6}{$-x$}
\psfrag{REP7}{$\tau$}
\psfrag{REP8}{$\sigma^\mathrm{z}(\tau,x_1)$}
\psfrag{REP9}{$\sigma^\mathrm{z}(0,x_2)$}
\psfrag{REP10}{$\Longleftrightarrow$}
\includegraphics[scale=0.18]{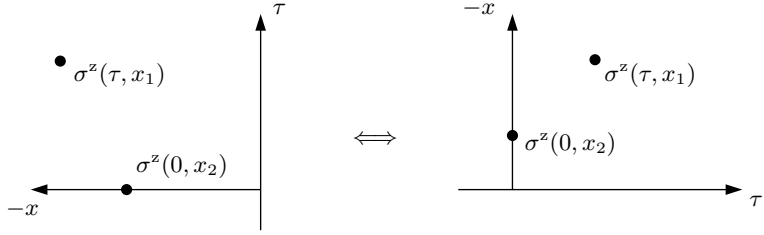}
\caption{Euclidean rotation. The boundary condition at $x=0$ turns into an 
  initial condition. The ``time'' in the rotated system runs
  between $x=0$ and $x=-\infty$.}
\end{figure}
\begin{figure}[t]
\begin{center}
\includegraphics[scale=0.3,clip=true]{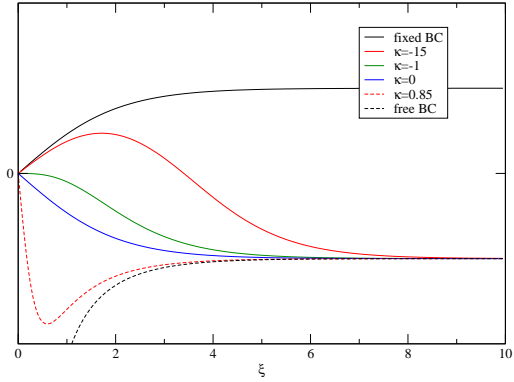}
\end{center}
\caption{$\hat{K}(\xi)$ as function of $\xi$. Note that 
  $h=2\sqrt{M}$ corresponds to $\kappa=-1$ and $h=h_\mathrm{c}=\sqrt{2M}$ to
  $\kappa=0$.}
\label{fig:khat}
\end{figure}

As we are working in the Euclidean formalism, $\tau$ and $x$ are
interchangeable.  
Therefore, one may also take $x$ to be the Euclidean time, which
implies that the equal-time section is the infinite line, $x=\text{const.}$,
$-\infty<\tau<\infty$, and the associated Hilbert space $\mathcal{H}$ is that
of the corresponding bulk theory \eqref{eq:defA}. The boundary at $x=0$ now
appears as an initial condition which is expressed in terms of a ``boundary
state'' $\ket{\mathrm{B}}$. As was shown by Ghoshal and
Zamolodchikov~\cite{GhoshalZamolodchikov94}, the correlation function
\eqref{eq:cfboundary} can then be expressed as
\begin{equation}
\label{eq:cfboundarystate}
C(\tau,x_1,x_2)=
\frac{\bra{0}\mathcal{T}_x\,\sigma^\mathrm{z}(\tau,x_1)\,
\sigma^\mathrm{z}(0,x_2)\ket{\mathrm{B}}}{\bra{0}\mathrm{B}\rangle}.
\end{equation}
Here $\mathcal{T}_x$ is the $x$-ordering operator, which orders the
largest $x_i$ to the right, and $\ket{0}\in\mathcal{H}$ is the ground
state of the model on the infinite line. The boundary state is 
given by
\begin{equation}
  \label{eq:boundarystate}
  \ket{\mathrm{B}}=\exp\left(\int_0^\infty\frac{d\xi}{2\pi}
    K(\xi)A^\dagger(-\xi)A^\dagger(\xi)\right)\ket{0},
\end{equation}
where $K(\xi)=R(\ii \pi/2-\xi)$. For the Ising model it is explicitly
given by
\begin{equation}
  \label{eq:defKh}
  K(\xi)=\ii \tanh\frac{\xi}{2}\ \frac{\kappa+\cosh\xi}{\kappa-\cosh\xi}.
\end{equation}
This simplifies to 
\begin{equation}
  \label{eq:defKfreefixed}
  K_\mathrm{free}(\xi)=-\ii \coth\frac{\xi}{2},\quad
  K_\mathrm{fixed}(\xi)=\ii \tanh\frac{\xi}{2},
\end{equation}
for free and fixed boundary conditions, respectively.  For later use we
introduce the real function $\hat{K}(\xi)=-\ii K(\xi)$, which is plotted for
several values of $\kappa$ in Fig.~\ref{fig:khat}.  We note that for $0\le
h\le 2\sqrt{M}$ the function $\hat{K}(\xi)$ is negative for all $\xi$, whereas
for $2\sqrt{M}<h$ the function $\hat{K}(\xi)$ is positive in a certain
$\xi$-interval.  Furthermore, for $0<h<h_\mathrm{c}=\sqrt{2M}$ one finds
$\hat{K}(\xi)<-1$ for certain $\xi$.  In this region the boundary scattering
matrix \eqref{eq:Rmatrix} has a pole in the strip
$0<\mathfrak{Im}\,\xi<\pi/2$, indicating the existence of a boundary bound
state~\cite{GhoshalZamolodchikov94}. We will discuss this state in more detail
in Sec.~\ref{sec:bbsexpl} below.  Finally, we note that for free boundary
conditions the boundary scattering matrix \eqref{eq:Rmatrix} has a pole at
$\xi=0$.  This results in the appearance of a zero-momentum mode in the
boundary state, \ie \eqref{eq:boundarystate} is replaced
by~\cite{GhoshalZamolodchikov94}
\begin{equation}
  \label{eq:boundarystatefree}
  \ket{\mathrm{B_{free}}}=\bigl(1+A^\dagger(0)\bigr)
  \exp\left(\int_0^\infty\frac{d\xi}{2\pi}
    K(\xi)A^\dagger(-\xi)A^\dagger(\xi)\right)\ket{0}.
\end{equation}
Since all form factors of $\sigma^\mathrm{z}$ involving an odd number of
particles vanish, this zero-momentum mode does not contribute to the two-point
function \eqref{eq:cfboundarystate} of $\sigma^\mathrm{z}$, and hence will be
ignored in our analysis. Finally, we note that $K(\xi)=-K(-\xi)$ and
$K(\xi)\rightarrow 0$ ($\xi\rightarrow 0$) except for free boundary
conditions.

As we have interchanged space and time and $x$ is running from $0$ to
$-\infty$ in the new framework, the $\tau$- and $x$-dependence of operators
$\sigma^\mathrm{z}(\tau,x)$ is now given by
\begin{equation}
\sigma^\mathrm{z}(\tau,x)=
e^{-xH}\,e^{-\ii\tau P}\,\sigma^\mathrm{z}(0,0)\,e^{\ii\tau P}\,e^{xH},
\end{equation}
where $H$ is the Hamiltonian of the system on the infinite line,
$-\infty<\tau<\infty$, and $P$ is the total momentum.

\subsection{Boundary bound state}\label{sec:bbsexpl}
If we consider free boundary conditions ($h=0$) there exist two degenerate
ground states $\ket{0_\mathrm{B},\pm}$. The effect of a small field
$0<h<h_\mathrm{c}=\sqrt{2M}$ is to split these into two non-degenerate states
$\ket{0_\mathrm{B}}$ and $\ket{1_\mathrm{B}}$, where $\ket{1_\mathrm{B}}$ can
be interpreted as a boundary bound state~\cite{GhoshalZamolodchikov94}. In
this domain we can parametrise $\kappa$ as
\begin{equation}
\label{eq:kappacos}
\kappa=\cos v,\quad 0<v<\frac{\pi}{2}.  
\end{equation}
The two states can be distinguished by the asymptotic behaviour of the
one-point function
$\bra{0_\mathrm{B}}\sigma^\mathrm{z}(x)\ket{0_\mathrm{B}}\rightarrow+\sigma_0$ 
or
$\bra{1_\mathrm{B}}\sigma^\mathrm{z}(x)\ket{1_\mathrm{B}}\rightarrow-\sigma_0$
for $x\rightarrow -\infty$, respectively (we assume that $h\ge 0$).  
The energy of
$\ket{1_\mathrm{B}}$ is given by $E_1=E_0+M\sin v$, where $E_0$ denotes the
ground-state energy. If $h$ approaches the critical value $h_\mathrm{c}$, the
boundary bound state becomes weakly bound and its effective size diverges. For
$h>h_\mathrm{c}$ \eqref{eq:Rmatrix} possesses no pole in the physical strip
and hence no boundary bound state occurs.

\begin{figure}[t]
\begin{center}
\includegraphics[scale=0.3,clip=true]{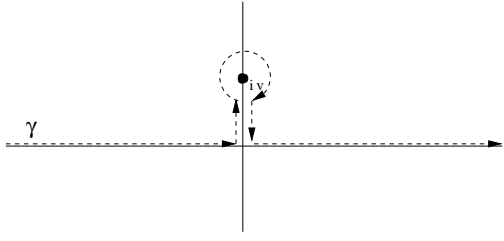}
\end{center}
\caption{Contour of integration for the excited boundary state.}
\label{fig:conexcitedbs}
\end{figure}

Correlation functions in the boundary bound state can be calculated following
Ref.~\onlinecite{GhoshalZamolodchikov94}
\begin{equation}
\bra{1_\mathrm{B}}\mathcal{T}_\tau\,\sigma^\mathrm{z}(\tau,x_1)\,
\sigma^\mathrm{z}(0,x_2)\ket{1_\mathrm{B}}=
\frac{\bra{0'}\mathcal{T}_x\,\sigma^\mathrm{z}(\tau,x_1)\,
\sigma^\mathrm{z}(0,x_2)\ket{\mathrm{B}'}}{\bra{0'}\mathrm{B}'\rangle},
\label{eq:cprime}
\end{equation}
where $\ket{0'}=\ket{0,-}\in\mathcal{H}$ denotes the ``wrong'' ground state.
The excited boundary state is given by~\cite{GhoshalZamolodchikov94}
\begin{equation}
  \label{eq:excitedboundarystate}
  \ket{\mathrm{B}'}=\exp\left(\frac{1}{2}\int_\gamma\frac{d\xi}{2\pi}
    K(\xi)A^\dagger(-\xi)A^\dagger(\xi)\right)\ket{0'},
\end{equation}
where the contour of integration is shown in Fig.~\ref{fig:conexcitedbs}.  The
contour encircles the pole of \eqref{eq:defKh} at $\xi=\ii v$, whose residue
equals $-2\ii\cot v\tan(v/2)$.

Due to the $\mathbb{Z}_2$-invariance of the Ising model in the bulk the matrix
elements of $\sigma^\mathrm{z}$ remain unchanged up to a minus sign when
considering the state $\ket{0'}$, \ie the form factors are
\begin{equation}
\label{eq:sigmaffprime}
\bra{0'}\sigma^\mathrm{z}\ket{\{\theta_{1},\ldots,\theta_{2n}\}'}\equiv
\bra{0'}\sigma^\mathrm{z}\,A^\dagger(\theta_{1})\ldots 
A^\dagger(\theta_{2n})\ket{0'}=
-\ii^n\sigma_0\prod_{\substack{i,j=1\\ i<j}}^{2n}
\tanh\frac{\theta_i-\theta_j}{2}.
\end{equation}
In particular, we have $E_0^\mathrm{bulk}=E_{0'}^\mathrm{bulk}$. The minus
sign can be deduced from
$\bra{0'}\sigma^\mathrm{z}\ket{0'}=-\bra{0}\sigma^\mathrm{z}\ket{0}$.  

\section{Green's functions}
In order to set the stage, we first calculate the Green's functions of
the Majorana fermions. The mode expansions in the presence of the
boundary and in the absence of a boundary bound state are given by 
\begin{eqnarray}
\psi(\tau,x)\!\!&=&\!\!\sqrt{\frac{M}{2}}
\int\limits_0^{\infty}\!\frac{d\theta}{\sqrt{2\pi}}\,
\biggl[c(\theta)\,e^{-\frac{\ii\pi}{4}}\,e^{-M\tau\cosh\theta}
\Bigl(e^{-\frac{\theta}{2}}\,e^{\ii Mx\sinh\theta}
+R(\theta)\,e^{\frac{\theta}{2}}\,e^{-\ii Mx\sinh\theta}\Bigr)\nonumber\\*
& &\hspace{20mm}+c^\dagger(\theta)\,e^{\frac{\ii\pi}{4}}\,e^{M\tau\cosh\theta}
\Bigl(e^{-\frac{\theta}{2}}\,e^{-\ii Mx\sinh\theta}
+R(-\theta)\,e^{\frac{\theta}{2}}\,e^{\ii Mx\sinh\theta}\Bigr)\biggr],
\label{eq:modepsiboundary}\\[2mm]
\bar{\psi}(\tau,x)\!\!&=&\!\!\sqrt{\frac{M}{2}}
\int\limits_0^{\infty}\!\frac{d\theta}{\sqrt{2\pi}}\,
\biggl[c(\theta)\,e^{\frac{\ii\pi}{4}}\,e^{-M\tau\cosh\theta}
\left(e^{\frac{\theta}{2}}\,e^{\ii Mx\sinh\theta}
+R(\theta)\,e^{-\frac{\theta}{2}}\,e^{-\ii Mx\sinh\theta}\right)\nonumber\\*
& &\hspace{20mm}+c^\dagger(\theta)\,e^{-\frac{\ii\pi}{4}}\,e^{M\tau\cosh\theta}
\Bigl(e^{\frac{\theta}{2}}\,e^{-\ii Mx\sinh\theta}
+R(-\theta)\,e^{-\frac{\theta}{2}}\,e^{\ii Mx\sinh\theta}\Bigr)\biggr].
\label{eq:modebarpsiboundary}
\end{eqnarray}
Here $c(\theta)$ and $c^\dagger(\theta)$ are canonical fermion
annihilation and creation operators
$\{c(\theta),c^\dagger(\theta')\}=2\pi\delta(\theta-\theta')$. 

\begin{figure}
\psfrag{REPa}{a)}
\psfrag{REPb}{b)}
\psfrag{REP11}{$\psi(0,x_2)$}
\psfrag{REP12}{oscillating behaviour}
\psfrag{REP13}{$\psi(0,x_2)$}
\hspace{-30mm}
\includegraphics[scale=0.18]{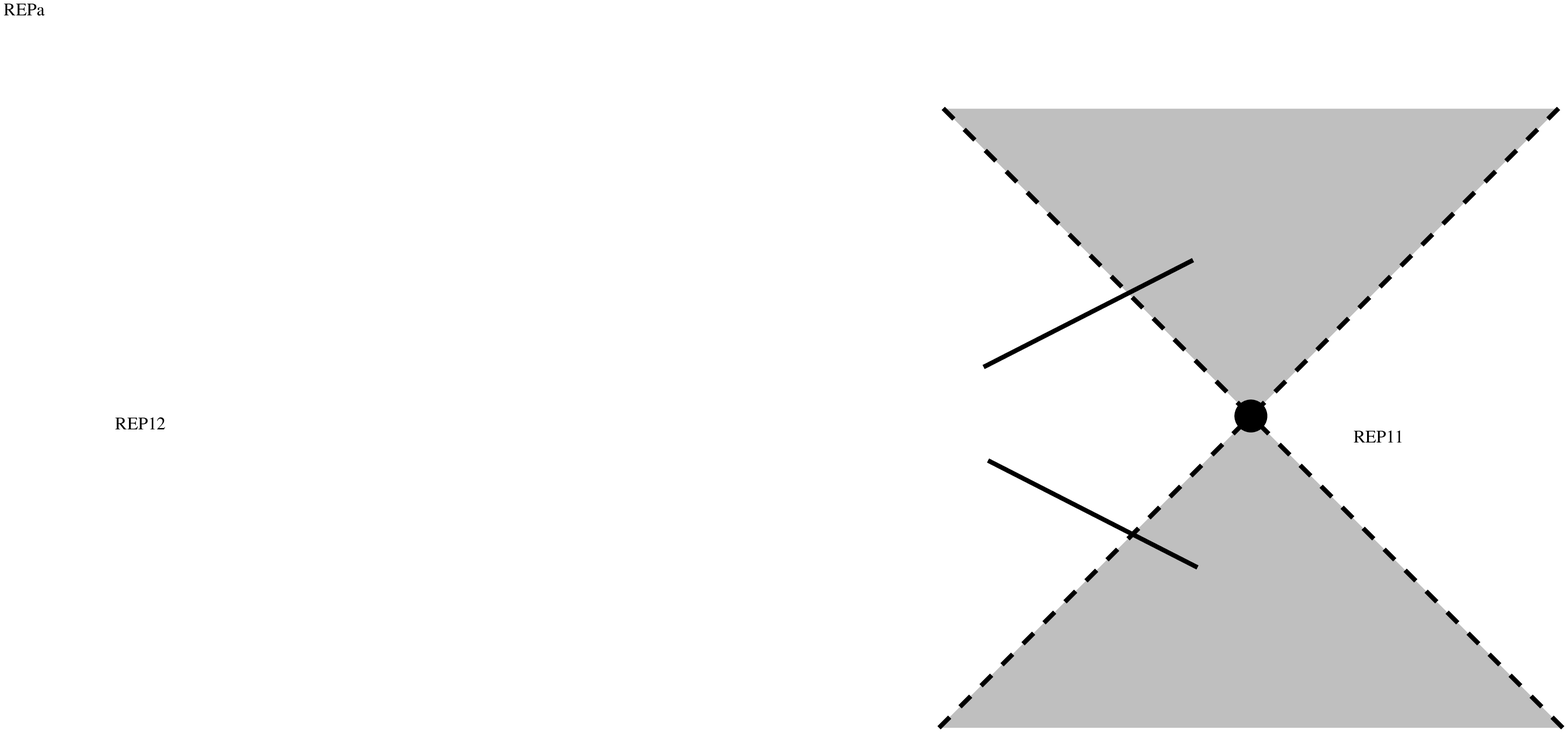}
\hspace{20mm}
\includegraphics[scale=0.18]{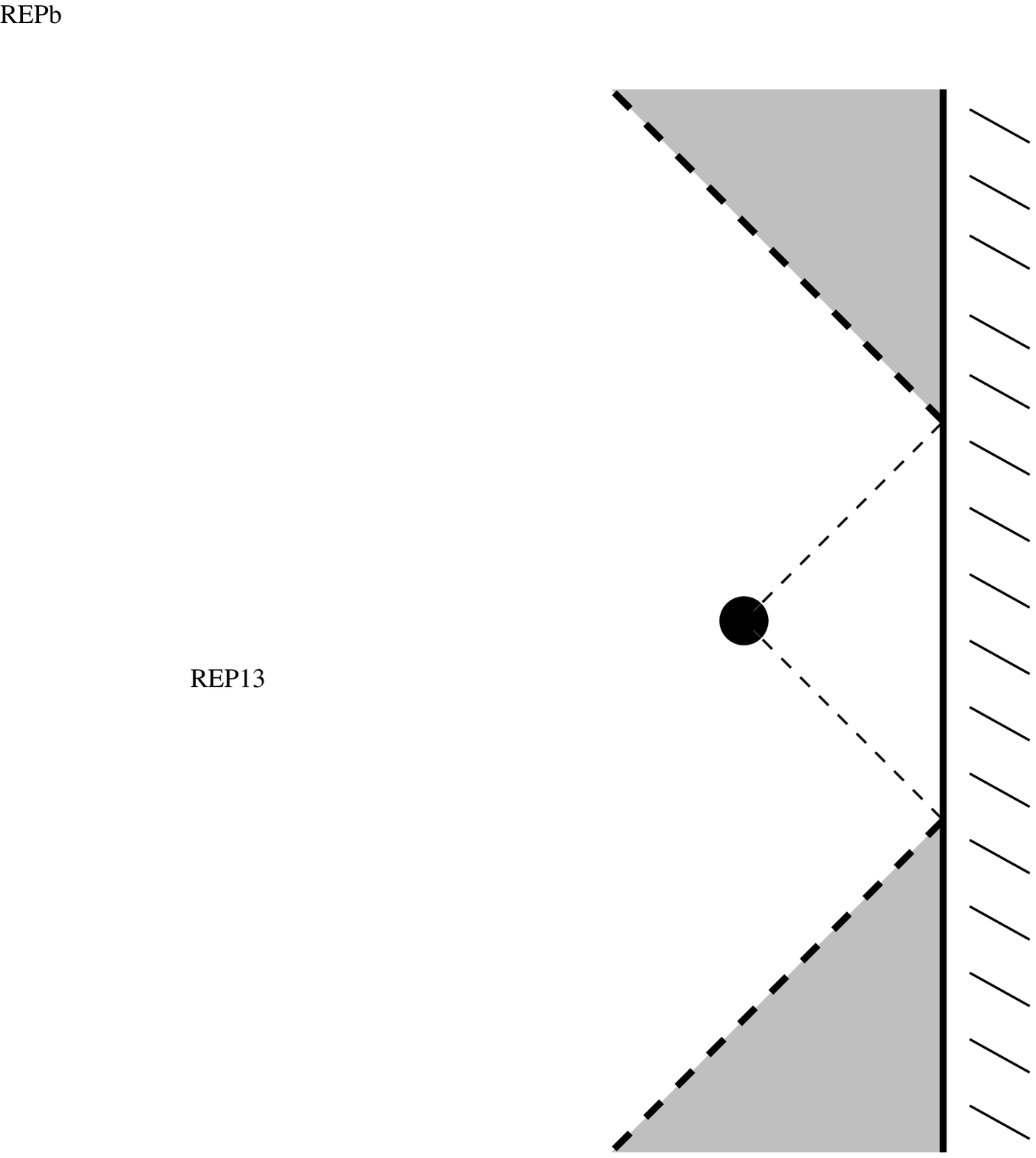}
\caption{Light-cone effects observed for the Green's function 
  \eqref{eq:GFpsi}. a) In the bulk one observes oscillating behaviour if
  $r^2<t^2$. b) The boundary contributions show oscillating behaviour if
  $4R^2<t^2$.}
\label{fig:lightcones}
\end{figure}

The Green's function of $\psi$ is now easily derived to be (we assume $\tau>0$
for simplicity)
\begin{equation}
\begin{split}
\bra{0_\mathrm{B}}\psi(\tau,x_1)\,\psi(0,x_2)\ket{0_\mathrm{B}}&=
\frac{M}{2}\int_{-\infty}^\infty d\theta\,e^{-M\tau\cosh\theta}\,
\Bigl[e^{-\theta}\,e^{\ii Mr\sinh\theta}+R(\theta)\,
e^{-2M\ii R\sinh\theta}\Bigr]\\*[2mm]
&=M\sqrt{\frac{\ii\tau-r}{\ii\tau+r}}\,K_1\bigl(M\sqrt{r^2+\tau^2}\bigr)
+\frac{M}{2}\int_{-\infty}^\infty d\theta\,R\bigl(\theta+\ii\theta_0\bigr)\,
e^{-M\sqrt{4R^2+\tau^2}\cosh\theta},
\end{split}
\label{eq:GFpsi}
\end{equation}
where $\theta_0=\arctan\bigl(2|R|/\tau\bigr)$ and $K_1$ denotes the modified
Bessel function of order one~\cite{AbramowitzStegun65}. Furthermore, we have
introduced centre-of-mass coordinates $R=(x_1+x_2)/2$ and $r=x_2-x_1$. We
stress that $R\le 0$. In real space, $\tau=\ii t$, the first term is
oscillating for $r^2<t^2$ and damped for $r^2>t^2$, \emph{i.e.} we
observe a light-cone effect. The second term is oscillating for
$4R^2<t^2$ and damped otherwise. The physical interpretation of the
oscillating behaviour is that for $4R^2<t^2$ a particle can propagate
from $(0,x_2)$ to $(t,x_1)$ via the boundary (see
Fig.~\ref{fig:lightcones}). If one calculates the Green's function
using the rotated system and the boundary state, one obtains an
additional phase of $\pi/2$. The physical origin of this phase is the
Lorentz spin $s_{\psi}=-1/2$ of $\psi$, which implies that the Green's
function transforms non-trivial under Lorentz rotations. In the case
$h<h_\mathrm{c}$ there is an additional term due to the presence of
the boundary bound state.  

In the same way one obtains (we assume $\tau>0$)
\begin{equation}
\bra{0_\mathrm{B}}\psi(\tau,x_1)\,\bar{\psi}(0,x_2)\ket{0_\mathrm{B}}
=-\ii M\,K_0\bigl(M\sqrt{r^2+\tau^2}\bigr)
-\ii\frac{M}{2}\sqrt{\frac{\ii\tau+2R}{\ii\tau-2R}}\,
\int_{-\infty}^\infty d\theta\,R\bigl(\theta+\ii\theta_0\bigr)\,
e^{\theta}\,e^{-M\sqrt{4R^2+\tau^2}\cosh\theta},
\end{equation}
which shows the same kind of oscillatory behaviour as \eqref{eq:GFpsi}.

\section{Local magnetisation}
In this section we briefly discuss the one-point function of the spin
operator. The one-point function in the ground state
$\ket{0_\mathrm{B}}$ was first calculated by Konik, LeClair and
Mussardo~\cite{Konik-96} and to second order in $K$ is given by
\begin{equation}
\begin{split}
\label{eq:op0}
\bra{0_\mathrm{B}}\sigma^\mathrm{z}(x)\ket{0_\mathrm{B}}&=
\sigma_0-\ii\sigma_0\int_0^\infty\frac{d\xi}{2\pi}\,
K(\xi)\,\tanh\xi\,e^{2Mx\cosh\xi}\\*[2mm]
&\hspace{4mm}-\frac{\sigma_0}{2}\int_0^\infty 
\frac{d\xi_1d\xi_2}{(2\pi)^2}\, 
\prod_{i=1}^2 K(\xi_i)\tanh\xi_i\,
\left(\frac{\cosh\xi_1-\cosh\xi_2}{\cosh\xi_1+\cosh\xi_2}\right)^2
e^{2Mx\sum_i\cosh\xi_i}+\ldots
\end{split}
\end{equation}
Here $x<0$ denotes the distance from the boundary. The full series can
be written as Fredholm determinant~\cite{Konik-96}.

\begin{figure}[t]
\begin{center}
\includegraphics[scale=0.3,clip=true]{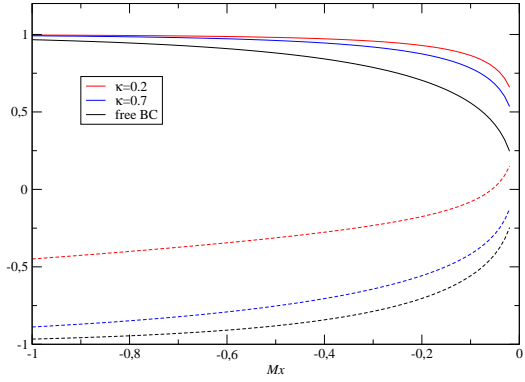}
\end{center}
\caption{Local magnetisation in the ground state and the
  boundary bound state for different values of the boundary magnetic
  field $h>0£$ up
  to first order in the boundary reflection matrix $K$.  The full lines
  represent the ground-state result \eqref{eq:op0}, the dashed lines the
  result for the boundary bound state \eqref{eq:op1}. }
\label{fig:opbbs}
\end{figure}

Using \eqref{eq:cprime}, \eqref{eq:excitedboundarystate} and
\eqref{eq:sigmaffprime}, we can calculate the expectation value of the
spin operator in the boundary bound state
\begin{equation}
\begin{split}
\bra{1_\mathrm{B}}\sigma^\mathrm{z}(x)\ket{1_\mathrm{B}}&=
-\sigma_0+\ii\frac{\sigma_0}{2}\int_\gamma\frac{d\xi}{2\pi}\,
K(\xi)\,\tanh\xi\,e^{2Mx\cosh\xi}+\ldots\\*[2mm]
&=-\sigma_0+\ii\sigma_0\int_0^\infty\frac{d\xi}{2\pi}\,
K(\xi)\,\tanh\xi\,e^{2Mx\cosh\xi}
+\sigma_0\,\tan\frac{v}{2}\,e^{2Mx\cos v}+\ldots,
\label{eq:op1}
\end{split}
\end{equation}
where the contour $\gamma$ is defined in Fig.~\ref{fig:conexcitedbs} and we
recall that $\kappa=1-h^2/2M=\cos v$.  We observe that
$\bra{1_\mathrm{B}}\sigma^\mathrm{z}(x)\ket{1_\mathrm{B}}\rightarrow
-\sigma_0$ for $x\rightarrow-\infty$ as expected. In the limit
$v\rightarrow\pi/2$ the last term remains comparable to $\sigma_0$ even for
large distances $x$ to the boundary. 
In this limit the boundary bound state becomes weakly
bound and its size diverges. The local magnetisations of the
ground state and the boundary bound state are shown in
Fig.~\ref{fig:opbbs}. For free
boundary conditions the two results equal each other up to a global minus
sign. For finite boundary magnetic field the spins in the vicinity of the
boundary are aligned parallel to $h$, which increases the local magnetisation.

\section{Spin-spin correlation function}\label{sec:corr}
We now turn to the calculation the leading terms in an expansion in
powers of the boundary reflection matrix $K$ of the two-point function
of Ising spins \eqref{eq:cfboundary}.  Using a spectral representation
of the correlator in terms of the scattering states \eqref{eq:defA} we
obtain the following expression
\begin{equation}
\label{eq:corr}
C(\tau,x_1,x_2)=\sum_{n=0}^\infty\sum_{m=0}^\infty C_{2n\,2m}(\tau,x_1,x_2),
\end{equation}
where
\begin{equation}
\begin{split}
C_{2n\,2m}(\tau,x_1,x_2)=&\frac{1}{m!}\frac{1}{(2n)!}
\int_0^\infty\frac{d\xi_1\ldots d\xi_m}{(2\pi)^m}
\int_{-\infty}^\infty\frac{d\theta_1\ldots d\theta_{2n}}{(2\pi)^{2n}}\,
K(\xi_1)\ldots K(\xi_m)\label{eq:corrmn}\\*[2mm]
&\hspace{5mm}\cdot\left\{
\begin{aligned}
\bra{0}\sigma^\mathrm{z}(\tau,x_1)\ket{\theta_{1},\ldots,\theta_{2n}}
\bra{\theta_{2n},\ldots,\theta_{1}}\sigma^\mathrm{z}(0,x_2)
\ket{-\xi_1,\xi_1,\ldots,-\xi_m,\xi_m}&,\; x_1<x_2,\\ 
\bra{0}\sigma^\mathrm{z}(0,x_2)\ket{\theta_{1},\ldots,\theta_{2n}}
\bra{\theta_{2n},\ldots,\theta_{1}}\sigma^\mathrm{z}(\tau,x_1)
\ket{-\xi_1,\xi_1,\ldots,-\xi_m,\xi_m}&, \; x_1>x_2.
\end{aligned}\right.
\end{split}
\end{equation}
We label the various terms in the double expansion (\ref{eq:corr})
by the numbers of particles in the intermediate state $2n$ and
in the boundary state $2m$, respectively. The connected correlator is
given by
\begin{equation}
\label{eq:corrconnected}
C_{\rm conn}(\tau,x_1,x_2)=
\bra{0_\mathrm{B}}\mathcal{T}_\tau\,\delta\sigma^\mathrm{z}(\tau,x_1)\,
\delta\sigma^\mathrm{z}(0,x_2)\ket{0_\mathrm{B}}=
C(\tau,x_1,x_2)-\bra{0_\mathrm{B}}\sigma^\mathrm{z}(x_1)\ket{0_\mathrm{B}}
\bra{0_\mathrm{B}}\sigma^\mathrm{z}(x_2)\ket{0_\mathrm{B}},
\end{equation}
where $\delta\sigma^\mathrm{z}(\tau,x)=\sigma^\mathrm{z}(\tau,x)-
\bra{0_\mathrm{B}}\sigma^\mathrm{z}(\tau,x)\ket{0_\mathrm{B}}$ and the second
term is in fact independent of $\tau$.

\subsection{Regularisation}
The functions \eqref{eq:corrmn} contain matrix elements of the form
\begin{equation}
\bra{\theta_1,\ldots,\theta_n}\sigma^\mathrm{z}\ket{\xi_1,\ldots,\xi_m},
\label{eq:generalme}
\end{equation}
which possess kinematical poles whenever $\theta_i=\xi_j$ and therefore need
to be regularised~\cite{Smirnov92book}. Let $A$ denote a set of
one-particle excitations and $A_1$ and $A_2$ a partition of $A$. 
The scattering matrix arising from the commutations
necessary to rewrite $\ket{A}$ as $\ket{A_2A_1}$ is denoted
by $S_{AA_1}$, \ie
$\ket{A}=S_{AA_1}\ket{A_2A_1}=S_{AA_2}\ket{A_1A_2}$. For example,
if $\ket{A}=\ket{\theta_1,\ldots,\theta_5}$ and
$\ket{A_1}=\ket{\theta_2,\theta_3}$, then  
\begin{equation}
\ket{\theta_1,\ldots,\theta_5}=
S_{AA_1}\ket{\theta_1,\theta_4,\theta_5,\theta_2,\theta_3}
\quad\text{with}\quad
S_{AA_1}=1.
\end{equation}
If $A$ and $B$ denote two sets of one-particle excitations, the regularisation
of the form factors \eqref{eq:generalme} reads~\cite{Smirnov92book}
\begin{eqnarray}
\bra{A}\sigma^\mathrm{z}\ket{B}&=&
\sum_{\substack{A=A_1 \cup A_2\\ B=B_1 \cup B_2}}
d(B_2)\,S_{AA_1}\,S_{B_1B}\,\bra{A_2}B_2\rangle\,
\bra{A_1+\ii 0}\sigma^\mathrm{z}\ket{B_1}\label{eq:reg1}\\*[2mm]
&=&\sum_{\substack{A=A_1 \cup A_2\\ B=B_1 \cup B_2}}
S_{AA_2}\,S_{B_2B}\,\bra{A_2}B_2\rangle\,
\bra{A_1-\ii 0}\sigma^\mathrm{z}\ket{B_1},\label{eq:reg2}
\end{eqnarray}
where the sums are over all possible ways to break the sets $A=A_1\cup A_2$
and $B=B_1\cup B_2$ into subsets.  The scalar products $\bra{A_2}B_2\rangle$
are easily evaluated using \eqref{eq:Aalgebra}. The factor $d(A)$ is
present by virtue of the semi-locality of the spin operator with
respect to the fundamental field and is given by
\begin{equation}
d(A)=(-1)^{n(A)},
\label{eq:defdA}
\end{equation}
where $n(A)$ denotes the number of elements in $A$. As all rapidities in the
remaining matrix elements are distinct, they can be evaluated using the
crossing relations
\begin{equation}
\begin{split}
\label{eq:connectedff}
\bra{\theta_{i_1}\!\pm\!\ii 0,\ldots,\theta_{i_p}\!\pm\!\ii 0}
\sigma^\mathrm{z}\ket{\xi_{j_1},\ldots,\xi_{j_q}}&=
\bra{0}\sigma^\mathrm{z}
\ket{\theta_{i_1}\!+\!\ii\pi\!\pm\!\ii \eta_{i_1},\ldots,
\theta_{i_p}\!+\!\ii\pi\!\pm\!\ii \eta_{i_p},
\xi_{j_1},\ldots,\xi_{j_q}}\\
&=f(\theta_{i_1}\!+\!\ii \pi\!\pm\!\ii \eta_{i_1},\ldots,
\theta_{i_p}\!+\!\ii \pi\!\pm\!\ii \eta_{i_p},\xi_{j_1},\ldots,\xi_{j_q}).
\end{split}
\end{equation}
We stress that equations \eqref{eq:reg1}, \eqref{eq:reg2} and
\eqref{eq:defdA} are valid only for the operator $\sigma^\mathrm{z}$
in the Ising model. For the disorder operator $\mu^\mathrm{z}$ we have
$l_{\mu^\mathrm{z}\mu^\mathrm{z}}=1$ and as a result the factor
$d(B_2)$ in \eqref{eq:reg1} needs to be dropped, whereas
\eqref{eq:reg2} remains unchanged.  In other theories and in
particular for operators with $l_{O\Psi}\neq\pm 1$ (where $\Psi$
denotes the fundamental field), additional phase factors related to
the non-locality of the operators appear.

We have checked the validity of \eqref{eq:reg1} and \eqref{eq:reg2} using a
finite-size regularisation for the Ising model (see below).

\subsection{Spin-spin correlation function in the bulk system}
\begin{figure}[t]
\begin{center}
\includegraphics[scale=0.3]{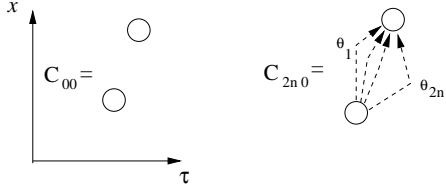}
\end{center}
\caption{Graphical representation of $C_{00}$ and $C_{2n\,0}$. The circles
  represent the operators $\sigma^\mathrm{z}$, where the upper one corresponds
  to the later time $x$. The particles created by the $A^\dagger(\theta)$'s
  are represented by the arrows. We order the arrows between the circles from
  left to right as $\theta_1,\ldots,\theta_{2n}$.}
\label{fig:diagrams1}
\end{figure}
The simplest terms in the expansion \eqref{eq:corr} are those with $m=0$, \ie
terms without boundary contributions.  We recall the definition of the
centre-of-mass coordinates $R=(x_1+x_2)/2\le 0$ and $r=x_2-x_1$. We 
find $C_{00}(\tau,r)=\sigma_0^2\propto M^{1/4}$ as well as
\begin{equation}
C_{2n\,0}(\tau,r)=\frac{1}{(2n)!}
\int_{-\infty}^\infty\frac{d\theta_1\ldots d\theta_{2n}}{(2\pi)^{2n}}\,
\big| f(\theta_1,\ldots,\theta_{2n})\big|^2\,
e^{-M|r|\sum_i\cosh\theta_i}\,e^{\ii \,\sgn{r}M\tau\sum_i\sinh\theta_i}.
\label{eq:C0n}
\end{equation}
We note that the exponential factor $e^{-M|r|\sum_i\cosh\theta_i}$ ensures the
convergence of the integrals. This is a general feature of the expressions we
obtain. As we will see later, the results for the spectral function remain
finite even in the limit $r\rightarrow 0$.  In order to facilitate the
comparison to the standard bulk results we now shift the contours of
integration in \eqref{eq:C0n} as follows. If we write $\theta=s+\ii \varphi$,
the exponential factors are given by
\begin{equation}
e^{-M|r|\cosh\theta}\,e^{\ii \,\sgn{r}M\tau\sinh\theta}=
e^{\ii (\sgn{r}M\tau\cos\varphi-M|r|\sin\varphi)\sinh s}
e^{-(\sgn{r}M\tau\sin\varphi+M|r|\cos\varphi)\cosh s},
\end{equation}
which vanishes exponentially for $s\rightarrow\pm\infty$ provided
that
\begin{equation}
\sgn{r}M\tau\sin\varphi+M|r|\cos\varphi>0.
\label{eq:condition}
\end{equation}
Hence, in the first and third quadrants 
of the $(\tau,r)$-plane, \ie when $\tau r>0$, the contributions at
$\mathfrak{Re}\,\theta_i=\pm\infty$ vanish as long as
$0\le\varphi\le\pi/2$ and we may
shift the contour of integration $\theta_i\rightarrow\theta_i+\ii \pi/2$
without changing the result. In the second and fourth quadrants 
($\tau r<0$) we may shift $\theta_i\rightarrow\theta_i-\ii \pi/2$
instead. As the form factors have no poles in the strip
$-\pi/2\le\mathfrak{Im}\,\theta_i\le\pi/2$, we obtain (see also
Appendix~\ref{sec:appuf}) 
\begin{equation}
C_{2n\,0}(\tau,r)=\frac{\sigma_0^2}{(2n)!}\int_{-\infty}^\infty 
\frac{d\theta_1\ldots d\theta_{2n}}{(2\pi)^{2n}}\,
\prod_{\substack{i,j=1\\ i<j}}^{2n}\tanh^2\frac{\theta_i-\theta_j}{2}\,
e^{-\ii M|r|\sum_i\sinh\theta_i}\,e^{-M|\tau|\sum_i\cosh\theta_i}.
\label{eq:Cn0change}
\end{equation}
In order to arrive a this expression we have replaced
$\theta_i\rightarrow-\theta_i$ in the second and fourth quadrants. The result
(\ref{eq:Cn0change}) is the well-known result for the spin-spin correlation
function in the scaling limit of the Ising model~\cite{ising1}. In comparison
to \eqref{eq:C0n} we have effectively reversed the Euclidean rotation done in
Sec.~\ref{sec:boundary}, \ie we have interchanged $\tau$ and $r$ back in order
to interpret $r$ as space and $\tau=\ii t$ as Euclidean time.

If we shift the contours of integration by
$\theta_i\rightarrow\theta_i\pm\ii\arctan\bigl(\tau/r\bigr)$, \eqref{eq:C0n}
can be cast in the following form
\begin{equation}
C_{2n\,0}(\tau,r)=\frac{\sigma_0^2}{(2n)!}
\int_{-\infty}^\infty\frac{d\theta_1\ldots d\theta_{2n}}{(2\pi)^{2n}}\,
\prod_{\substack{i,j=1\\ i<j}}^{2n}\tanh^2\frac{\theta_i-\theta_j}{2}\,
e^{-M\sqrt{r^2+\tau^2}\sum_i\cosh\theta_i}.
\label{eq:C0nuniversal}
\end{equation}
In real space, $\tau=\ii t$, we observe oscillating behaviour for time-like
separations and damped behaviour for space-like separations.

\subsection{First-order contributions in the boundary reflection matrix}
\begin{figure}[t]
\begin{center}
  \includegraphics[scale=0.3]{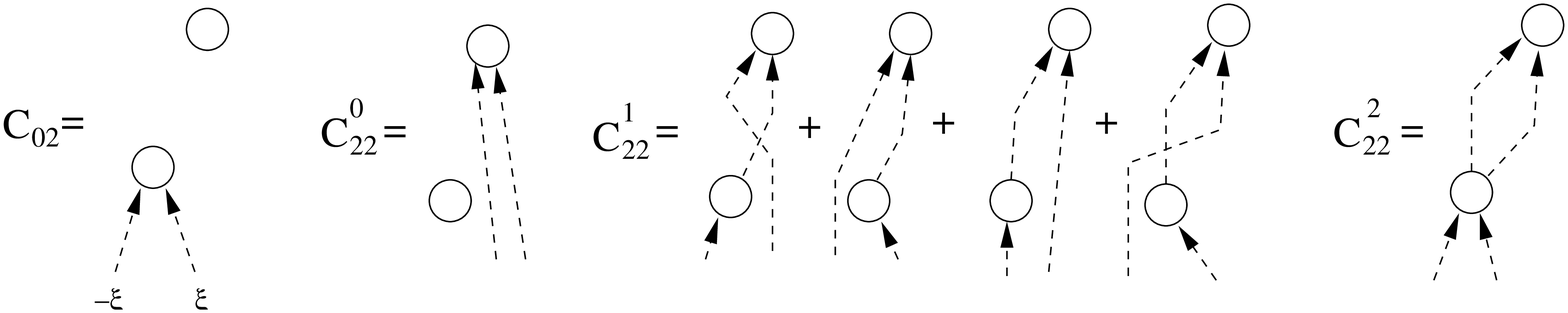}
\end{center}
\caption{Graphical representation of $C_{02}$ and $C_{22}$. The arrows 
  starting at the lower edge represent the particles $A^\dagger(-\xi)$ and
  $A^\dagger(\xi)$ coming from the boundary state. If these arrows pass the
  lower circle, which represents an operator $\sigma^\mathrm{z}$, it indicates
  that some of the internal particles created by the $A^\dagger(\theta_i)$'s
  have been contracted with the external lines, \ie terms like
  $\bra{\theta_i}\xi\rangle=2\pi\delta(\theta_i-\xi)$ appear in the
  corresponding formulas. The upper index denotes the number of lines
  connecting the two operators.}
\label{fig:diagrams2}
\end{figure}
In this subsection we calculate the leading contribution in the boundary
reflection matrix $K$. The first, time-independent, term is given by
\begin{equation}
C_{02}(\tau,x_1,x_2)=-\ii \sigma_0^2\int_0^\infty\frac{d\xi}{2\pi}\,K(\xi)\,
\tanh\xi\,e^{2M\max(x_1,x_2)\cosh\xi}.
\label{eq:C01}
\end{equation}
We note that $C_{02}$ is cancelled in the connected correlation function
\eqref{eq:corrconnected}.
The first term containing a contribution of the boundary to the two-particle
continuum is also the first term which contains matrix elements of the form
\eqref{eq:generalme},
\begin{equation}
\begin{split}
\label{eq:C11first}
C_{22}(\tau,x_1,x_2)=&\frac{1}{2}\int_0^\infty\frac{d\xi}{2\pi}
\int_{-\infty}^\infty\frac{d\theta_1 d\theta_2}{(2\pi)^2}\,
K(\xi)\,f(\theta_1,\theta_2)\,
\bra{\theta_2,\theta_1}\sigma^\mathrm{z}\ket{-\xi,\xi}\\*
&\hspace{10mm}\cdot
e^{2M\max(x_1,x_2)\cosh\xi}\,
e^{-M|r|\sum_i\cosh\theta_i}\,e^{\ii\sgn{r}M\tau\sum_i\sinh\theta_i}.
\end{split}
\end{equation}
For the evaluation of the second matrix element we can use either
\eqref{eq:reg1} or \eqref{eq:reg2}, which give
\begin{eqnarray}
\bra{\theta_2,\theta_1}\sigma^\mathrm{z}\ket{-\xi,\xi}
&=&\phantom{-}(2\pi)^2\,
\sigma_0\bigl(\delta(\theta_1+\xi)\,\delta(\theta_2-\xi)-
\delta(\theta_1-\xi)\,\delta(\theta_2+\xi)\bigr)\nonumber\\*
& &+2\pi\delta(\theta_1-\xi)\, 
f(\theta_2+\ii \pi+\ii \eta_2,-\xi)
-2\pi\delta(\theta_1+\xi)\, 
f(\theta_2+\ii \pi+\ii \eta_2,\xi)\nonumber\\*
& &-2\pi\delta(\theta_2-\xi)\, 
f(\theta_1+\ii \pi+\ii \eta_1,-\xi)
+2\pi\delta(\theta_2+\xi)\, 
f(\theta_1+\ii \pi+\ii \eta_1,\xi)\nonumber\\*
& &+f(\theta_2+\ii \pi+\ii \eta_2,\theta_1+\ii \pi+\ii \eta_1,
-\xi,\xi)\label{eq:C11reg1}\\[2mm]
&=&\phantom{-}(2\pi)^2\,
\sigma_0\bigl(\delta(\theta_1+\xi)\,\delta(\theta_2-\xi)-
\delta(\theta_1-\xi)\,\delta(\theta_2+\xi)\bigr)\nonumber\\*
& &-2\pi\delta(\theta_1-\xi)\, 
f(\theta_2+\ii \pi-\ii \eta_2,-\xi)
+2\pi\delta(\theta_1+\xi)\, 
f(\theta_2+\ii \pi-\ii \eta_2,\xi)\nonumber\\*
& &+2\pi\delta(\theta_2-\xi)\, 
f(\theta_1+\ii \pi-\ii \eta_1,-\xi)
-2\pi\delta(\theta_2+\xi)\, 
f(\theta_1+\ii \pi-\ii \eta_1,\xi)\nonumber\\*
& &+f(\theta_2+\ii \pi-\ii \eta_2,\theta_1+\ii \pi-\ii \eta_1,
-\xi,\xi).\label{eq:C11reg2}
\end{eqnarray}
Both regularisations give altogether six contributions, which can be
represented graphically as shown in Fig.~\ref{fig:diagrams2}. 
Either regularisation scheme gives rise
to a contribution
\begin{equation}
C_{22}^0=
-\ii \sigma_0^2\int_0^\infty\frac{d\xi}{2\pi}K(\xi)\,\tanh\xi\,
e^{2M\min(x_1,x_2)\cosh\xi},
\end{equation}
where the additional upper index denotes the number of lines
connecting the two operators. The contribution $C_{22}^0$ is
similar to $C_{02}$ and like the latter cancels in the connected
correlation function. 

The next term, $C_{22}^1$, actually consists of four contributions which can
be combined using the relation $K(-\xi)=-K(\xi)$
\begin{equation}
C_{22}^1=\mp\sigma_0^2\int_{-\infty}^\infty
\frac{d\xi}{2\pi}\frac{d\theta}{2\pi}\,K(\xi)\,
\tanh\frac{\xi-\theta}{2}\,\coth\frac{\xi+\theta\pm\ii \eta}{2}\,
e^{2MR\cosh\xi}\,e^{\ii \,\sgn{r}M\tau\sinh\xi}\,
e^{-M|r|\cosh\theta}\,e^{\ii \,\sgn{r}M\tau\sinh\theta}.
\label{eq:C112}
\end{equation}
For free boundary conditions the integral over $\xi$ has to be understood as
a principal value integration. Furthermore, the upper sign corresponds to the
regularisation \eqref{eq:C11reg1} and the lower sign to \eqref{eq:C11reg2}.
The difference between the two is compensated by analogous differences
in the contribution $C_{22}^2$ discussed below.  In order to take the limit
$\eta\rightarrow 0$ we shift the contour of the $\theta$-integration in the
same way as we did for the bulk terms $C_{2n\,0}$. As the exponential factors
containing $\theta$ in \eqref{eq:C112} equal those in \eqref{eq:C0n}, we
obtain again the condition \eqref{eq:condition} for the vanishing of the
integrals at $\mathfrak{Re}\,\theta\rightarrow\pm\infty$. Hence, if the
space-time coordinates $(\tau,r)$ lie in the first or third quadrants, $\tau
r>0$, we have to shift $\theta\rightarrow\theta+\ii\pi/2$. In order to avoid
the appearance of extra terms arising from the residues of the coth in
\eqref{eq:C112} we choose the regularisation \eqref{eq:C11reg1}, for
which the pole in the $\theta$-plane is located at
$\theta=-\xi-\ii\eta$, \ie below the real axis.  By the same reasoning
we choose the regularisation \eqref{eq:C11reg2} if $\tau r<0$.  Doing
so we obtain  
\begin{eqnarray}
C_{22}^1&=&\mp\sigma_0^2\int_{-\infty}^\infty
\frac{d\xi}{2\pi}\frac{d\theta}{2\pi}\,K(\xi)\,
\tanh\frac{\xi-\theta\mp\ii\pi/2}{2}\,
\coth\frac{\xi+\theta\pm\ii\pi/2\pm\ii\eta}{2}\nonumber\\*
& &\hspace{20mm}\cdot
e^{2MR\cosh\xi}\,e^{\ii\sgn{r}M\tau\sinh\xi}\,
e^{-\ii\sgn{\tau}Mr\sinh\theta}\,e^{-M|\tau|\cosh\theta}\\[2mm]
&=&\pm\sigma_0^2\int_{-\infty}^\infty\frac{d\xi}{2\pi}
\frac{d\theta}{2\pi}\,K(\xi)\,
\frac{\cosh\theta\pm\ii\sinh\xi}{\cosh\theta\mp\ii\sinh\xi}\,
e^{2MR\cosh\xi}\,e^{-\ii\sgn{\tau}Mr\sinh\theta}\,
e^{\ii\sgn{r}M\tau\sinh\xi}\,e^{-M|\tau|\cosh\theta},
\label{eq:C1122}
\end{eqnarray}
where we have taken the limit $\eta\rightarrow 0$. In the second and
fourth quadrants of the $(\tau,r)$-plane we now change variables
$\xi\rightarrow -\xi$ and $\theta\rightarrow -\theta$, which
in particular changes $e^{\ii\sgn{r}M\tau\sinh\xi}$ to $e^{\ii
  M|\tau|\sinh\xi}$. Hence, as $R<0$ we can shift $\xi\rightarrow
\xi+\ii\pi/2$ and obtain 
\begin{eqnarray}
C_{22}^1(\tau,x_1,x_2)&=&-\sigma_0^2\int_{-\infty}^\infty\frac{d\xi}{2\pi}
\frac{d\theta}{2\pi}\,K\bigl(\xi+\ii\tfrac{\pi}{2}\bigr)\,
\frac{\cosh\xi-\cosh\theta}{\cosh\xi+\cosh\theta}\,
e^{\ii 2MR\sinh\xi}\,e^{-\ii M|r|\sinh\theta}\,
e^{-M|\tau|(\cosh\theta+\cosh\xi)}
\label{eq:C112osc}\\*[2mm]
& &+\Theta(h_{\rm c}-h)\ 2\sigma_0^2\,\cot v\,\tan\frac{v}{2}\,e^{2MR\cos v}\,
e^{-M|\tau|\sin v}\,\nonumber\\*
& &\hspace{20mm}\cdot
\int_{-\infty}^\infty\frac{d\theta}{2\pi}\,
\frac{\cosh\theta-\sin v}{\cosh\theta+\sin v}\,
e^{-\ii M|r|\sinh\theta}\,e^{-M|\tau|\cosh\theta}.
\label{eq:C112bbs}
\end{eqnarray}
Here \eqref{eq:C112bbs} originates from the pole of $K(\xi)$ at $\xi=\ii v$
(we recall $\kappa=1-h^2/2M=\cos v$) and is present if
$h<h_\mathrm{c}$. In particular, for free boundary conditions we
obtain $\sigma_0^2\,e^{2MR}\,K_0\bigl(M\sqrt{r^2+\tau^2}\bigr)/\pi$.
Explicit expressions for $K\bigl(\xi+\ii\tfrac{\pi}{2}\bigr)$ are given in
\eqref{eq:defKhpi2}--\eqref{eq:defKfixedpi2}. We will see below that
\eqref{eq:C112osc} yields an oscillating contribution to the spectral
function. 

Going through the same steps as above, we find that the final term in
the regularisation \eqref{eq:C11reg1} or \eqref{eq:C11reg2}
respectively can be cast in the form
\begin{eqnarray}
C_{22}^2&=&-\ii \frac{\sigma_0^2}{2}\int_0^\infty\frac{d\xi}{2\pi}
\int_{-\infty}^\infty\frac{d\theta_1d\theta_2}{(2\pi)^2}\,K(\xi)\,
\tanh\xi\,\prod_{i=1}^2\coth\frac{\xi-\theta_i\mp\ii\eta_i}{2}\,
\coth\frac{\xi+\theta_i\pm\ii\eta_i}{2}\nonumber\\*
& &\hspace{20mm}\cdot
\tanh^2\frac{\theta_1-\theta_2}{2}\,
e^{2M\max(x_1,x_2)\cosh\xi}\,e^{-M|r|\sum_i\cosh\theta_i}\,
e^{\ii\sgn{r}M\tau\sum_i\sinh\theta_i}\\[2mm]
&=&-\ii \frac{\sigma_0^2}{2}\int_0^\infty\frac{d\xi}{2\pi}
\int_{-\infty}^\infty\frac{d\theta_1d\theta_2}{(2\pi)^2}\,K(\xi)\,
\tanh\xi\,\tanh^2\frac{\theta_1-\theta_2}{2}\prod_{i=1}^2
\frac{\cosh\xi\pm\ii \sinh\theta_i}{\cosh\xi\mp\ii \sinh\theta_i}\nonumber\\*
& &\hspace{20mm}\cdot
e^{2M\max(x_1,x_2)\cosh\xi}\,e^{\mp\ii M|r|\sum_i\sinh\theta_i}\,
e^{-M|\tau|\sum_i\cosh\theta_i}.
\label{eq:C113}
\end{eqnarray}
Here we have shifted $\theta_i\rightarrow\theta_i+\ii \pi/2$ for $\tau r>0$
and $\theta_i\rightarrow\theta_i-\ii \pi/2$ for $\tau r<0$, respectively, and
taken the limits $\eta_i\rightarrow 0$.  Finally, we can substitute
$\theta_i\rightarrow -\theta_i$ if $\tau r<0$. 

Puting everything together, the result for $C_{22}$ reads
\begin{eqnarray}
C_{22}(\tau,x_1,x_2)&=&
-\ii\sigma_0^2\int_0^\infty\frac{d\xi}{2\pi}K(\xi)\,\tanh\xi\,
e^{2M\min(x_1,x_2)\cosh\xi}\nonumber\\[2mm]
& &-\sigma_0^2\int_{-\infty}^\infty\frac{d\xi}{2\pi}
\frac{d\theta}{2\pi}\,K\bigl(\xi+\ii\tfrac{\pi}{2}\bigr)\,
\frac{\cosh\xi-\cosh\theta}{\cosh\xi+\cosh\theta}\,
e^{\ii 2MR\sinh\xi}\,e^{-\ii M|r|\sinh\theta}\,
e^{-M|\tau|(\cosh\theta+\cosh\xi)}\nonumber\\[2mm]
& &-\ii \frac{\sigma_0^2}{2}\int_0^\infty\frac{d\xi}{2\pi}
\int_{-\infty}^\infty\frac{d\theta_1d\theta_2}{(2\pi)^2} K(\xi)\,
\tanh\xi\,\tanh^2\frac{\theta_1-\theta_2}{2}\prod_{i=1}^2
\frac{\cosh\xi+\ii \sinh\theta_i}{\cosh\xi-\ii \sinh\theta_i}\label{eq:C_11}\\*
& &\hspace{20mm}\cdot 
e^{2M\max(x_1,x_2)\cosh\xi}\,
e^{-\ii M|r|\sum_i\sinh\theta_i}\,e^{-M|\tau|\sum_i\cosh\theta_i}
\nonumber\\[2mm]
& &+\Theta(h_{\rm c}-h)\ 2\sigma_0^2\,\cot
v\,\tan\frac{v}{2}\,e^{2MR\cos v}\, 
e^{-M|\tau|\sin v}\nonumber\\*
& &\hspace{20mm}\cdot
\int_{-\infty}^\infty\frac{d\theta}{2\pi}\,
\frac{\cosh\theta-\sin v}{\cosh\theta+\sin v}\,
e^{-\ii M|r|\sinh\theta}\,e^{-M|\tau|\cosh\theta}.\nonumber
\end{eqnarray}
The last term is the contribution due to the boundary bound state.

If one uses regularisation \eqref{eq:C11reg2} instead of \eqref{eq:C11reg1} in
the first or third quadrants of the $(\tau,r)$-plane, one has to keep track of
the poles of the coth in \eqref{eq:C1122} as well as \eqref{eq:C113} in the
strip $0\le\mathfrak{Im}\,\theta\le\pi/2$. A straightforward calculation shows
that the contributions of these residues cancel out and that one finds again
the result \eqref{eq:C_11}. In order to check the regularisation scheme, we
have performed a finite-size regularisation of $C_{22}$. The results are
presented in Appendix~\ref{sec:finite} and equal \eqref{eq:C_11} within a
relative error of less than $10^{-4}$.

In order to study the light-cone effect in more detail, let us reconsider
$C_{22}^1$ in the region $h_\mathrm{c}<h$. We start with \eqref{eq:C1122} and
shift the contours of integration $\theta\rightarrow\theta-\ii\theta_0$ and
$\xi\rightarrow\xi+\ii\xi_0$, where $\theta_0=\arctan\bigl(|r|/|\tau|\bigr)$
and $\xi_0=\arctan\bigl(|\tau|/2|R|\bigr)$. This yields
\begin{equation}
C_{22}^1
\!=\sigma_0^2\int_{-\infty}^\infty\frac{d\xi}{2\pi}
\frac{d\theta}{2\pi}\,K\bigl(\xi+\ii\xi_0\bigr)\,
\frac{\cosh\bigl(\theta\!-\!\ii\theta_0\bigr)\!+
\!\ii\sinh\bigl(\xi\!+\!\ii\xi_0\bigr)}
{\cosh\bigl(\theta\!-\!\ii\theta_0\bigr)\!-
\!\ii\sinh\bigl(\xi\!+\!\ii\xi_0\bigr)}\,
e^{-M\sqrt{r^2+\tau^2}\cosh\theta}\,e^{-M\sqrt{4R^2+\tau^2}\cosh\xi}.
\end{equation}
We observe the same qualitative features as for the Green's function
\eqref{eq:GFpsi}, \ie oscillating behaviour for $r^2<t^2$ as well as
$4R^2<t^2$.

In Appendix~\ref{sec:appcorr} we calculate further contributions to first
order in the boundary reflection matrix, but with higher numbers of particles
in the intermediate state. In particular, we show that
$C_{42}^2(\tau,x_1,x_2)=C_{22}^2(\tau,x_2,x_1)$.

\subsection{Second-order contributions in the boundary reflection matrix}
\label{sec:Cmge2}
In this subsection we calculate the leading contributions to second order in
the boundary reflection matrix $K$. The first term of this kind, $C_{04}$,
drops out in the calculation of the connected correlation function. The next
term is $C_{24}$.  For simplicity, we will restrict ourselves to the first
quadrant in the $(\tau,r)$-plane, \ie $\tau>0$ and $r=x_2-x_1>0$:
\begin{equation}
\begin{split}
C_{24}(\tau,x_1,x_2)=&\;\frac{1}{4}\int_0^\infty\frac{d\xi_1d\xi_2}{(2\pi)^2}
\int_{-\infty}^\infty\frac{d\theta_1 d\theta_2}{(2\pi)^2}\,
K(\xi_1)\,K(\xi_2)\,e^{2Mx_2\sum_i\cosh\xi_i}\,
e^{-Mr\sum_i\cosh\theta_i}\\[2mm]
&\hspace{15mm}\cdot e^{\ii M\tau\sum_i\sinh\theta_i}\,f(\theta_1,\theta_2)\,
\bra{\theta_2,\theta_1}\sigma^\mathrm{z}\ket{-\xi_1,\xi_1,-\xi_2,\xi_2}.
\end{split}
\label{eq:C12}
\end{equation}
\begin{figure}[t]
\begin{center}
\includegraphics[scale=0.3]{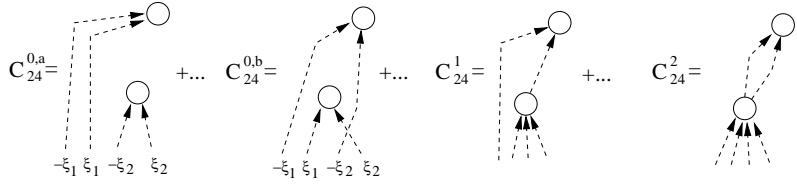}
\end{center}
\caption{Graphical representation of $C_{24}$. The first term, 
  $C_{24}^{0,\mathrm{a}}$, is completely disconnected. In the second term,
  $C_{24}^{0,\mathrm{b}}$, the contractions of the particles in the
  intermediate state and the particles in the boundary state are crossed and
  the operators become intertwined.}
\label{fig:diagrams5}
\end{figure}
Using \eqref{eq:reg1} the second matrix element can be written as
\begin{equation}
\begin{split}
\bra{\theta_2,\theta_1}\sigma^\mathrm{z}\ket{-\xi_1,\xi_1,-\xi_2,\xi_2}
=&\;\bra{\theta_2,\theta_1}-\xi_1,\xi_1\rangle\,f(-\xi_2,\xi_2)+
\bra{\theta_2,\theta_1}-\xi_2,\xi_2\rangle\,f(-\xi_1,\xi_1)\\
&-\bra{\theta_2,\theta_1}\xi_1,\xi_2\rangle\,f(-\xi_1,-\xi_2)
+\bra{\theta_2,\theta_1}-\xi_1,\xi_2\rangle\,f(\xi_1,-\xi_2)\\
&+\bra{\theta_2,\theta_1}\xi_1,-\xi_2\rangle\,f(-\xi_1,\xi_2)
-\bra{\theta_2,\theta_1}-\xi_1,-\xi_2\rangle\,f(\xi_1,\xi_2)\\
&-\bra{\theta_2}\xi_2\rangle\,
f(\theta_1+\ii \pi+\ii \eta_1,-\xi_1,\xi_1,-\xi_2)
+\bra{\theta_2}-\xi_2\rangle\,
f(\theta_1+\ii \pi+\ii \eta_1,-\xi_1,\xi_1,\xi_2)\\
&-\bra{\theta_2}\xi_1\rangle\,
f(\theta_1+\ii \pi+\ii \eta_1,-\xi_1,-\xi_2,\xi_2)
+\bra{\theta_2}-\xi_1\rangle\,
f(\theta_1+\ii \pi+\ii \eta_1,\xi_1,-\xi_2,\xi_2)\\
&+\bra{\theta_1}\xi_2\rangle\,
f(\theta_2+\ii \pi+\ii \eta_2,-\xi_1,\xi_1,-\xi_2)
-\bra{\theta_1}-\xi_2\rangle\,
f(\theta_2+\ii \pi+\ii \eta_2,-\xi_1,\xi_1,\xi_2)\\
&+\bra{\theta_1}\xi_1\rangle\,
f(\theta_2+\ii \pi+\ii \eta_2,-\xi_1,-\xi_2,\xi_2)
-\bra{\theta_1}-\xi_1\rangle\,
f(\theta_2+\ii \pi+\ii \eta_2,\xi_1,-\xi_2,\xi_2)\\
&+f(\theta_2+\ii \pi+\ii \eta_2,\theta_1+\ii \pi+\ii \eta_1,
-\xi_1,\xi_1,-\xi_2,\xi_2).
\end{split}
\label{eq:regC12}
\end{equation}
A graphical representation for the contribution \eqref{eq:C12} using
the decomposition \eqref{eq:regC12} is shown in
Fig.~\ref{fig:diagrams5}. We observe that there exist two 
different terms  in which the operators are not directly
connected. The first 
one, $C_{24}^{0,\mathrm{a}}$, is completely disconnected
and hence does not contribute to the connected correlation function
\eqref{eq:corrconnected}.  The second one,
$C_{24}^{0,\mathrm{b}}$, is obtained from the second and third lines of
\eqref{eq:regC12}. Although there is no line connecting the two
operators, the contractions of the particles in the intermediate state and the
particles in the boundary state are crossed and the operators become
intertwined. Hence, this term does contribute to the connected correlation
function. It is explicitly given by (for free boundary conditions the
integrals have to be understood as principal value integrations)
\begin{eqnarray}
C_{24}^{0,\mathrm{b}}(\tau,x_1,x_2)&=&
-\frac{\sigma_0^2}{2}\int_{-\infty}^\infty
\frac{d\xi_1d\xi_2}{(2\pi)^2}\,
K(\xi_1)\,K(\xi_2)\,\tanh^2\frac{\xi_1-\xi_2}{2}\,
e^{2MR\sum_i\cosh\xi_i}\,e^{\ii M\tau\sum_i\sinh\xi_i}\nonumber\\[2mm]
&=&-\frac{\sigma_0^2}{2}\int_{-\infty}^\infty
\frac{d\xi_1d\xi_2}{(2\pi)^2}\,
K\bigl(\xi_1+\ii\tfrac{\pi}{2}\bigr)\,K\bigl(\xi_2+\ii\tfrac{\pi}{2}\bigr)
\tanh^2\frac{\xi_1-\xi_2}{2}\,
e^{\ii 2MR\sum_i\sinh\xi_i}\,e^{-M\tau\sum_i\cosh\xi_i}
\label{eq:C122osc}\\*[2mm]
& &-\Theta(h_{\rm c}-h)\ 2\sigma_0^2\,\cot v\,\tan\frac{v}{2}\,e^{2MR\cos v}\,
e^{-M|\tau|\sin v}\nonumber\\*
& &\hspace{20mm}\cdot
\int_{-\infty}^\infty\frac{d\xi}{2\pi}\,K\bigl(\xi+\ii\tfrac{\pi}{2}\bigr)\,
\tanh^2\frac{\xi+\ii(\pi/2-v)}{2}\,
e^{\ii 2MR\sinh\xi}\,e^{-M\tau\cosh\xi}.\label{eq:C122bbs}
\end{eqnarray}
We note that the result for free boundary conditions is obtained as
the limit $v\rightarrow 0$.  In the derivation of \eqref{eq:C122bbs}
we used that the function had no pole at $\xi_1=\xi_2=\ii v$. Contribution
\eqref{eq:C122osc} can be rewritten as
\begin{equation}
-\frac{\sigma_0^2}{2}\int_{-\infty}^\infty
\frac{d\xi_1d\xi_2}{(2\pi)^2}\,
K(\xi_1+\ii\xi_0)\,K(\xi_2+\ii\xi_0)\,\tanh^2\frac{\xi_1-\xi_2}{2}\,
e^{-M\sqrt{4R^2+\tau^2}\sum_i\cosh\xi_i},
\end{equation}
which again shows oscillating behaviour for $4R^2<t^2$ in real time.  The
other terms appearing in \eqref{eq:regC12} yield sub-leading contributions to
the correlation function and are determined in Appendix~\ref{sec:appcorr}.

The calculation of higher-order terms in the boundary reflection matrix can be
performed along the same lines as above (see Appendix~\ref{sec:appcorr}).
Although no principal problems appear, the calculations become soon
rather tedious. The expectation is, however, that the first few orders in $K$
will give accurate results even rather close to the boundary. We 
show below that this is indeed the case for the local spectral
function, where the terms calculated above are found to be sufficient
for $M|R|\agt 0.2$.  The calculation of terms in the spectral representation
with a higher number of particles in the intermediate state is also
discussed in Appendix~\ref{sec:appcorr}. The corresponding terms in
the local spectral function are found to be negligible compared to the
two-particle contributions. The analogous behaviour for the bulk
Ising model is well
documented~\cite{YurovZamolodchikov91,DelfinoMussardo95,EsslerKonik05},
and it was argued by Cardy and Mussardo that this behaviour is a
general feature of form factor expansions in integrable field
theories~\cite{CardyMussardo93}.  

\subsection{Limiting case}
Let us consider the behaviour of the two-point function in the limit
$R\rightarrow-\infty$ in more detail. We have
$\min(x_1,x_2)\rightarrow-\infty$, but $\max(x_1,x_2)$ may remain
close to the boundary. Hence, in addition to oscillating terms like
$C_{22}^1$ there are contributions of terms like $C_{22}^2$,
$C_{42}^4$ and $C_{24}^2$. These terms diverge in the limit
$\max(x_1,x_2)\rightarrow 0$, \ie the series expansion \eqref{eq:corr}
ceases to converge. In the case $\min(x_1,x_2)=0$ one has to deal
with an operator located at the boundary. The properties of 
such boundary operators and their counterparts in the bulk are, in general,
very different.  For example, $\sigma^\mathrm{z}$ has the boundary scaling
dimension $1/2$, while its scaling dimensions in the bulk are
$(1/16,1/16)$~\cite{Cardy89}.  The treatment of boundary operators in the
framework of integrable field theories and form factor expansions has been put
forward recently by Bajnok, Palla and Tak\'{a}cs~\cite{Bajnok-06}.

The scaling behaviour of the two-point function in the case of free
boundary conditions was derived for the lattice Ising model by
Bariev~\cite{Bariev}.  He obtained the following result in
the region $\sqrt{r^2+\tau^2}>2R\gg 1/M$, 
\begin{equation}
\label{eq:Bariev1}
\bra{0_\mathrm{B}}\delta\sigma^\mathrm{z}(\tau,x_1)\, 
\delta\sigma^\mathrm{z}(0,x_2)\ket{0_\mathrm{B}}\sim
\frac{1}{(r^2+\tau^2)^{1/4}}\,e^{2MR}\,e^{-M\sqrt{r^2+\tau^2}}.
\end{equation}
In the field theoretical calculation the leading term of the left-hand side
equals $C_{22}^1$, which is according to \eqref{eq:C112osc} and
\eqref{eq:C112bbs} given by
\begin{eqnarray}
C_{22}^1(\tau,x_1,x_2)&=&
\frac{\sigma_0^2}{\pi}\,e^{2MR}\,K_0\bigl(M\sqrt{r^2+\tau^2}\bigr)
\label{eq:limitK}\\*
& &-\sigma_0^2\int_{-\infty}^\infty\frac{d\xi}{2\pi}
\frac{d\theta}{2\pi}\,K\bigl(\xi+\ii\tfrac{\pi}{2}\bigr)\,
\frac{\cosh\xi-\cosh\theta}{\cosh\xi+\cosh\theta}\,
e^{\ii 2MR\sinh\xi}\,e^{-\ii M|r|\sinh\theta}\,
e^{-M|\tau|(\cosh\theta+\cosh\xi)}.
\label{eq:limit2}
\end{eqnarray}
In the region $\sqrt{r^2+\tau^2}\gg 1/M$ the asymptotic
behaviour~\cite{AbramowitzStegun65} of the first term is given by
\eqref{eq:Bariev1}. On the other hand we can evaluate the second term in the
stationary phase approximation, which yields
\begin{equation}
-\frac{\sigma_0^2}{2\pi M}\,K\bigl(\xi_0+\ii\tfrac{\pi}{2}\bigr)\,
\frac{\sqrt{r^2+\tau^2}-\sqrt{4R^2+\tau^2}}
{\sqrt{r^2+\tau^2}+\sqrt{4R^2+\tau^2}}\,
\frac{e^{-M\sqrt{4R^2+\tau^2}}}{(4R^2+\tau^2)^{1/4}}\,
\frac{e^{-M\sqrt{r^2+\tau^2}}}{(r^2+\tau^2)^{1/4}},
\quad\xi_0=\ii\arctan\left(\frac{2R}{|\tau|}\right).
\label{eq:limit3}
\end{equation}
The condition $\sqrt{r^2+\tau^2}>2R\gg 1/M$ translates into
$\tau^2>4x_1x_2>0$, where the second inequality follows from the fact that the
result for $C_{22}^1$ is valid only if both operators are located away from
the boundary.  The condition $\tau^2>0$, however, implies that
\eqref{eq:limit3} is exponentially suppressed compared to \eqref{eq:limitK}.
Hence the asymptotic behaviour of $C_{22}^1$ is given by \eqref{eq:limitK} and
equals the result derived by Bariev in the lattice Ising model.

We would also like to comment on the conformal limit $M\rightarrow 0$ of the
Ising model. The boundary conditions compatible with conformal symmetry are
the fixed and free ones. Using the method of mirror
images~\cite{Cardy84,DiFrancescoMathieuSenechal97} one can derive the
two-point function of $\sigma^\mathrm{z}$.  If one considers the limit
$x_2\rightarrow 0$ while keeping $x_1$ and $\tau$ fixed, one finds
\begin{equation}
\bra{0_\mathrm{B}}\sigma^\mathrm{z}(1/M,1/M)\,
\sigma^\mathrm{z}(0,x_2)\ket{0_\mathrm{B}}\sim
\left\{
\begin{aligned}
x_2^{3/8}&,\quad\text{free boundary conditions},\\
x_2^{-1/8}&,\quad\text{fixed boundary conditions}.\\
\end{aligned}\right.
\label{eq:cftres2}
\end{equation}
Although we cannot apply the expansion \eqref{eq:corr} in this limit, we
mention that this result is in agreement with the leading correction to the
correlation function derived above. For $\min(x_1,x_2)$ and $\tau$ kept fixed
this correction is given by \eqref{eq:C01}, which is negative in the case of
free boundary conditions and hence suggesting the vanishing of the correlation
function, but positive for fixed boundary conditions, thus supporting a
diverging behaviour.  However, in order to study the vicinity of the
conformally invariant point more accurately one has use different methods like
the truncated conformal space approach~\cite{TCSA,Dorey-98,Dorey-01}.

Finally, we mention that it is not possible to resum higher-order
contributions in $K$ by a geometric series, which can be achieved for
finite-temperature correlation functions in the Ising
model~\cite{Leclair-96npb} and the non-linear sigma model~\cite{Konik03}.

\section{Spectral function}\label{sec:specralfunction}
In order to gain further physical insight in the spin correlations it
is useful to calculate the corresponding spectral function.  
The positive frequency part is given by
\begin{equation}
C(\omega,x_1,x_2)=
\int_0^\infty d\tau\,e^{\ii \bar{\omega}\tau}\,C(\tau,x_1,x_2)
\biggl|_{\bar{\omega}\rightarrow-\ii\omega+\delta},
\label{eq:FTdef}
\end{equation}
where the analytic continuation of the frequencies is $\bar{\omega}=-\ii
\omega+\delta$.

There are essentially four different types of terms in the series
\eqref{eq:corrconnected}, which we will discuss below separately. First, we
recover the known results for the bulk from all terms with $m=0$. Second,
there exist terms which essentially yield corrections to the two- and
four-particle continua already present in the bulk. The leading
corrections of this kind are $C_{22}^2$ and $C_{42}^2$ as well as
$C_{42}^4$ and $C_{62}^4$. Third, we find oscillatory behaviour, which
is present even deep in the bulk ($M|R|\gg 1$).  Here the leading
contributions are  $C_{22}^1$ and $C_{24}^{0,\mathrm{b}}$.  Fourth, we
discuss the contributions of the boundary bound state for boundary
magnetic fields $h<h_\mathrm{c}$.  The numerical evaluation of the
integrals appearing in the spectral functions was performed using the
VEGAS routine for Monte Carlo
integration~\cite{PressTeukolskyVetterlingFlannery06}. 

\subsection{Bulk Result}
As we have mentioned before, the bulk results for the $n$-particle
level are given by $C_{2n\,0}$.  The full correlation function in the
bulk is 
\begin{equation}
C_\mathrm{bulk}(\tau,r)=\sum_{n=0}^\infty\frac{\sigma_0^2}{(2n)!}
\int_{-\infty}^\infty\frac{d\theta_1\ldots d\theta_{2n}}{(2\pi)^{2n}}\,
\prod_{\substack{i,j=1\\ i<j}}^{2n}\tanh^2\frac{\theta_i-\theta_j}{2}\,
e^{-\ii M|r|\sum_i\sinh\theta_i}\,e^{-M|\tau|\sum_i\cosh\theta_i}.
\label{eq:Cbulk}
\end{equation}
If we perform the Fourier transformation $\tau\rightarrow \omega$ and
calculate the spectral function
$S_\mathrm{bulk}=\mathfrak{Im}\,C_{\mathrm{bulk}}$, we find for the
two-particle continuum
\begin{equation}
S^2_\mathrm{bulk}(\omega,r)=
\frac{\sigma_0^2\pi}{2}\int_{-\infty}^\infty\frac{d\theta_1d\theta_2}{(2\pi)^2}
\tanh^2\frac{\theta_1-\theta_2}{2}\,
\cos\Bigl(Mr\sum\nolimits_i\sinh\theta_i\Bigr)\,
\delta\Bigl(\omega-M\sum\nolimits_i\cosh\theta_i\Bigr),
\label{eq:s2bulk}
\end{equation}
where we have restricted ourselves to positive frequencies $\omega>0$.
Resolving the $\delta$-functions gives
\begin{equation}
S^2_\mathrm{bulk}(\omega,r)=2\sigma_0^2
\int\limits_{0}^{\text{Arcosh}\bigl(\tfrac{\omega}{2M}\bigr)}
\frac{d\theta}{2\pi}
\frac{\tanh^2\theta}{\sqrt{\omega^2-4 M^2 \cosh^2\theta}}\,
\,\cos\Bigl(|r|\sqrt{\omega^2-4 M^2 \cosh^2\theta}\Bigr).
\label{eq:2bulk}
\end{equation}
We stress that \eqref{eq:2bulk} vanishes very slowly for large separation $r$
at fixed frequency $\omega$. A similar behaviour is found for the two-point
function of the disorder operator $\mu^\mathrm{z}$, whose leading
contribution is given by the one-particle peak (see
Sec.\ref{sec:disordercorr}).  The four-particle continuum can be
readily calculated from the corresponding term in \eqref{eq:Cbulk}.
One finds that the ratio $S^4_\mathrm{bulk}/S^2_\mathrm{bulk}$ is
$\sim 1/150$ at $\omega/M=25$ and smaller for lower energies. This
suppression of the higher-order terms in the series \eqref{eq:Cbulk}
is a well-known feature in the Ising model and other massive
theories~\cite{YurovZamolodchikov91,CardyMussardo93,DelfinoMussardo95,
  DelfinoCardy98,Controzzi-01,EsslerKonik05}.

\subsection{Two- and four-particle continua}
We next turn to terms in the spectral representation \eqref{eq:corr}, 
which have a direct counterpart in the bulk \eqref{eq:Cbulk}.

The two-particle continuum to first order in the boundary reflection matrix is
given by $C_{22}^2$ and $C_{42}^2$, \ie terms with two lines connecting
the two operators. Performing the Fourier transform, analytically
continuing and taking the imaginary part we find after a straightforward
calculation using the formulas given in Appendix~\ref{sec:appuf} 
\begin{equation}
\begin{split}
S^2_{1K}(\omega,R,r)&=
\sigma_0^2\int_0^\infty\frac{d\xi}{2\pi}
\int_{-\theta'}^{\theta'}\frac{d\theta}{2\pi}\,
\frac{\hat{K}(\xi)\,\tanh\xi}{\sqrt{(\omega-M\cosh\theta)^2-M^2}}\,
\frac{e^{2MR\cosh\xi}}{\cosh^2\xi+\sinh^2\theta}\,
\frac{\tanh^2\frac{\theta-\tilde{\theta}}{2}}
{\cosh^2\xi+\sinh^2\tilde{\theta}}\\*[2mm]
&\hspace{20mm}\cdot\biggl\{
\Bigl[\bigl(\cosh^2\xi-\sinh^2\theta\bigr)
\bigl(\cosh^2\xi-\sinh^2\tilde{\theta}\bigr)
-4\cosh^2\xi\sinh\theta\sinh\tilde{\theta}\Bigr]\\
&\hspace{65mm}\cdot\cos\bigl(M|r|(\sinh\theta+\sinh\tilde{\theta})\bigr)\,
\cosh\bigl(M|r|\cosh\xi\bigr)\\[2mm]
&\hspace{25mm}+
2\cosh\xi\,
\Bigl[\bigl(\cosh^2\xi-\sinh^2\tilde{\theta}\bigr)\sinh\theta
+(\cosh^2\xi-\sinh^2\theta\bigr)\sinh\tilde{\theta}\Bigr]\\
&\hspace{65mm}\cdot\sin\bigl(M|r|(\sinh\theta+\sinh\tilde{\theta})\bigr)\,
\sinh\bigl(M|r|\cosh\xi\bigr)\biggr\},
\end{split}
\label{eq:S1K2}
\end{equation}
where 
\begin{equation}
\label{eq:delta1}
\tilde{\theta}=\tilde{\theta}(\omega,\theta)=
\text{Arcosh}\biggl(\frac{\omega}{M}-\cosh\theta\biggr)
\quad\text{and}\quad
\theta'=\text{Arcosh}\biggl(\frac{\omega}{M}-1\biggr).
\end{equation}
We observe that the integrand is exponentially suppressed for large distances
from the boundary by the factor $e^{2MR\cosh\xi}$.  Furthermore, we note that
\eqref{eq:S1K2} is particle-hole symmetric. The two-particle continuum to
second order in $K$, which we denote by $S_{2K}^2$, is obtained from
$C_{24}^2$, $C_{44}^{2,\mathrm{a}}$ and $C_{64}^2$. Its explicit form is
derived in Appendix~\ref{sec:appS2K}.

\begin{figure}[t]
\begin{center}
  \includegraphics[scale=0.3,clip=true]{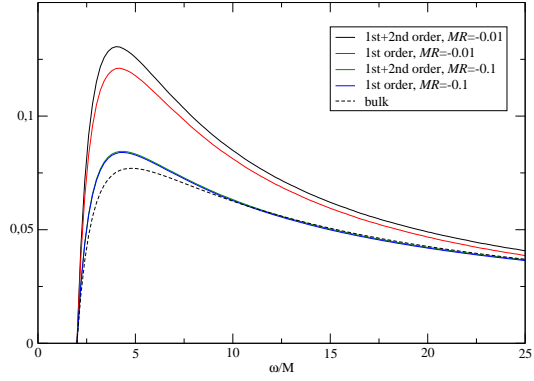}
\end{center}
\caption{Total two-particle contribution to the local spectral
  density \eqref{eq:Stot2K} for fixed boundary conditions (in units of
  $\sigma^2_0/M$). The bulk result is obtained for $MR\rightarrow-\infty$. We
  observe that for $MR=-0.1$ the second order contribution is already
  very small.}
\label{fig:totalfixed}
\end{figure}
\begin{figure}[t]
\begin{center}
\includegraphics[scale=0.3,clip=true]{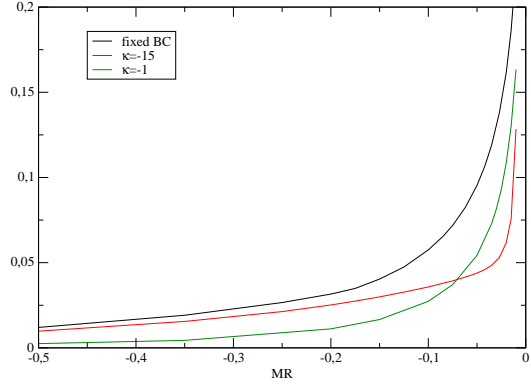}
\end{center}
\caption{Ratio of the first and second order (in the boundary
  reflection matrix) contributions to the two-particle continua 
  $S_{2K}^2/S_{1K}^2$ at their respective peaks for different
  distances from the boundary.} 
\label{fig:ratio}
\end{figure}

The total two-particle contribution to the local spectral density is given by
summation of all terms with two lines connecting the two operators. In
Fig.~\ref{fig:totalfixed} we have plotted the truncation at the second order
in $K$,
\begin{equation}
\label{eq:Stot2K}
S_\mathrm{tot}^2(\omega,R,r=0)=S^2_\mathrm{bulk}(\omega,r=0)+
S_{1K}^2(\omega,R,r=0)+S_{2K}^2(\omega,R,r=0)+\ldots,
\end{equation}
for fixed boundary conditions. The results show that the effects of the
boundary can be treated in perturbation theory for $M|R|\agt 0.2$. This is
also supported by the study of the ratio of $S_{2K}^2$ to $S_{1K}^2$ at their
respective peaks, which is plotted in Fig.~\ref{fig:ratio}. For example, for
fixed boundary conditions and $MR=-0.1$ the latter is about $5 \%$.

Starting from $C_{42}^4$ and $C_{62}^4$ we have calculated the four-particle
contribution to the spectral function $S_{1K}^4$ to first order in $K$.  The
explicit result is given in Appendix~\ref{sec:appS2K}. Comparison with
$S_{1K}^2$ shows that the four-particle contribution is negligible in the
low-energy region $\omega\alt 25M$.  This indicates that the fast rate of
convergence in the number of particles in the intermediate state, which is a
well-known fact for the Ising model in the
bulk~\cite{YurovZamolodchikov91,CardyMussardo93,DelfinoMussardo95,EsslerKonik05},
is also present in the system with boundary.

\subsection{Oscillating contributions}
The corrections to the bulk we have studied so far are negligible
for $MR\ll -1$. In this section we discuss oscillating terms which affect
the spectral function also deep in the bulk. The leading term of this kind is
given by \eqref{eq:C112osc}, we will denote it by $C_{22}^{1,\mathrm{osc}}$.
The corresponding contribution to the spectral function reads
\begin{equation}
\begin{split}
\label{eq:osc1}
S_{22}^{1,\mathrm{osc}}(\omega,R,r=0)=&\,\frac{2\sigma_0^2}{\omega}
\int\limits_0^{\text{Arcosh}\left(\frac{\omega}{M}-1\right)}\frac{d\xi}{2\pi}\,
\frac{\omega-2M\cosh\xi}{\sqrt{(\omega-M\cosh\xi)^2-M^2}}\\*[2mm]
&\hspace{5mm}\cdot
\Bigl[\mathfrak{Re}K\bigl(\xi+\ii\tfrac{\pi}{2}\bigr)\,
\cos\bigl(2MR\sinh\xi\bigr)-
\mathfrak{Im}K\bigl(\xi+\ii\tfrac{\pi}{2}\bigr)\,\sin\bigl(2MR\sinh\xi\bigr)
\Bigr],
\end{split}
\end{equation}
where $K\bigl(\xi+\ii\tfrac{\pi}{2}\bigr)$ is given by
\eqref{eq:defKhpi2}--\eqref{eq:defKfixedpi2}.  For simplicity we have
restricted ourselves to $r=0$. We stress that \eqref{eq:osc1} is not
exponentially suppressed for large distances from the boundary. In fact, the
integral \eqref{eq:osc1} possesses a square-root singularity at the upper
limit $\xi=\text{Arcosh}\left(\frac{\omega}{M}-1\right)$. Expanding the
non-singular part of the integrand about this value yields
\begin{equation}
  S_{22}^{1,\mathrm{osc}}(\omega,R,r=0)
  \sim\frac{\cos\bigl(\Delta R+\varphi\bigr)}{\sqrt{M|R|}},
    \quad\Delta=2\sqrt{\omega^2-2M\omega},
    \label{eq:osc1sim}
\end{equation}
where $\varphi$ depends on the boundary conditions. 

The next term we want to analyse is $C_{24}^{0,\mathrm{b},\mathrm{osc}}$. The
corresponding spectral function is found to be
\begin{equation}
\label{eq:osc2}
\begin{split}
S_{24}^{0,\mathrm{b},\mathrm{osc}}(\omega,R,r)=&-\frac{\sigma_0^2}{2M}
\int_{-\xi'}^{\xi'}\frac{d\xi}{2\pi}\,
\frac{1}{\sqrt{(\omega-M\cosh\xi)^2-M^2}}\,
\tanh^2\frac{\xi-\tilde{\xi}}{2}\\*[2mm]
&\hspace{0mm}\cdot
\biggl\{
\Bigl[\mathfrak{Re}K\bigl(\xi+\ii\tfrac{\pi}{2}\bigr)\,
\mathfrak{Re}K\bigl(\tilde{\xi}+\ii\tfrac{\pi}{2}\bigr)
-\mathfrak{Im}K\bigl(\xi+\ii\tfrac{\pi}{2}\bigr)\,
\mathfrak{Im}K\bigl(\tilde{\xi}+\ii\tfrac{\pi}{2}\bigr)\Bigr]\,
\cos\bigl(2MR(\sinh\xi+\sinh\tilde{\xi})\bigr)\\*[2mm]
&\hspace{18mm}-2\,\mathfrak{Re}K\bigl(\xi+\ii\tfrac{\pi}{2}\bigr)\,
\mathfrak{Im}K\bigl(\tilde{\xi}+\ii\tfrac{\pi}{2}\bigr)\,
\sin\bigl(2MR(\sinh\xi+\sinh\tilde{\xi})\bigr)\biggr\},
\end{split}
\end{equation}
where $\tilde{\xi}$ and $\xi'$ are given by
\begin{equation}
\tilde{\xi}=\tilde{\xi}(\omega,\xi)=
\text{Arcosh}\left(\frac{\omega}{M}-\cosh\xi\right),\quad
\xi'=\text{Arcosh}\left(\frac{\omega}{M}-1\right).
\label{eq:defdeltaxi}
\end{equation}
We note that \eqref{eq:osc2} is independent of $r$.  Numerical evaluation of
the integral in the range $MR\le -20$ and $\omega\approx 5M$ shows that
\begin{equation}
  S_{24}^{0,\mathrm{b},\mathrm{osc}}(\omega,R,r=0)
  \sim\frac{\cos\bigl(\Delta' R+\varphi'\bigr)}{(M|R|)^{3/2}},
    \quad\Delta'>\Delta.
    \label{eq:osc2sim}
\end{equation}
The larger oscillation frequency $\Delta'$ is due to the additional term
$\sinh\tilde{\xi}\ge 0$ in the sine and cosine in \eqref{eq:osc2}.  Although
\eqref{eq:osc2sim} falls off faster than \eqref{eq:osc1sim},
$S_{24}^{0,\mathrm{b},\mathrm{osc}}$ should not be interpreted as a
higher-order correction to $S_{22}^{1,\mathrm{osc}}$, as both terms are
present deep in the bulk. The analysis of the corresponding contributions in
the presence of a boundary bound state (see below) rather suggests that
$S_{24}^{0,\mathrm{b},\mathrm{osc}}$ represents an independent oscillating
contribution of order $K$; and hence that the total oscillating contribution
to first order in the boundary reflection matrix is given by
\begin{equation}
S_{1K}^\mathrm{osc}(\omega,R,r)=S_{22}^{1,\mathrm{osc}}(\omega,R,r)+
S_{24}^{0,\mathrm{b},\mathrm{osc}}(\omega,R,r).
\label{eq:osc1K}
\end{equation}
The higher-order corrections in $K$ to \eqref{eq:osc1} and \eqref{eq:osc2} are
in fact obtained starting from
$C_{24}^{1,\mathrm{osc}}+C_{44}^{1,\mathrm{osc}}$ and
$C_{26}^{0,\mathrm{osc}}+C_{46}^{0,\mathrm{osc}}$, respectively. The
corresponding spectral function $S_{2K}^\mathrm{osc}$ is evaluated in
Appendix~\ref{sec:apposc}.  We note that the appearing integrands are
exponentially suppressed for large distances from the boundary.

\begin{figure}[t]
\begin{center}
\includegraphics[scale=0.3,clip=true]{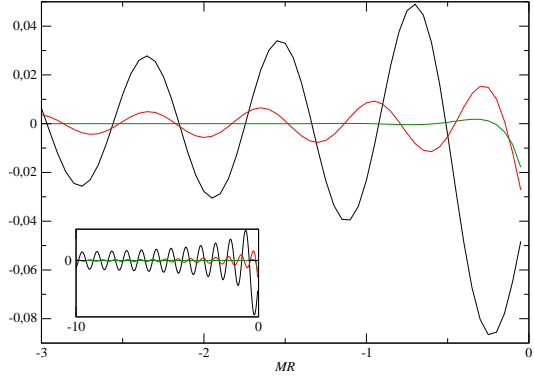}
\end{center}
\caption{Oscillating terms $S_{22}^{1,\mathrm{osc}}$ (black line) and 
  $S_{24}^{0,\mathrm{b},\mathrm{osc}}$ (red line) as well as
  $S_{24}^{1,\mathrm{osc}}+S_{44}^{1,\mathrm{osc}}$ (green line) for fixed
  boundary conditions, $\omega=5M$ and $r=0$ (in units of $\sigma^2_0/M$). We
  stress that $S_{22}^{1,\mathrm{osc}}$ and
  $S_{24}^{0,\mathrm{b},\mathrm{osc}}$ are present deep in the bulk, whereas
  $S_{24}^{1,\mathrm{osc}}+S_{44}^{1,\mathrm{osc}}$ is strongly suppressed for
  large $M|R|$.}
\label{fig:friedel}
\end{figure}
We have plotted \eqref{eq:osc1} and \eqref{eq:osc2} as well as \eqref{eq:osc3}
in Fig.~\ref{fig:friedel} for fixed boundary conditions, fixed energy and
varying distance from the boundary. The first two are present in the bulk
($MR\sim -10$), whereas the third one is strongly suppressed for large $M|R|$.
This underlines the interpretation of \eqref{eq:osc3} as first correction to
\eqref{eq:osc1}.  Finally, the correction to \eqref{eq:osc1} with three
particles in the intermediate state is given by $C_{42}^{3,\mathrm{osc}}$. It
is found to be negligible for all values of $MR$ and energies $\omega\le 25M$.
The same result holds for $C_{44}^{2,\mathrm{b},\mathrm{osc}}$, the first
correction to \eqref{eq:osc2} with more internal lines.  This again underlines
that the series \eqref{eq:corr} is rapidly convergent as $n\rightarrow\infty$.

\subsection{Contributions from the boundary bound state}
In the previous section we have investigated the oscillatory behaviour of the
two-point function. This behaviour was caused by the terms $C_{22}^1$,
$C_{24}^{0,\mathrm{b}}$, $C_{24}^1+C_{44}^1$ and $C_{26}^0+C_{46}^0$. In
Sec.~\ref{sec:corr} we have seen that in the range $h<h_\mathrm{c}$ all these
terms possess additional contributions due to the existence of the boundary
bound state, which we will now discuss in detail.

We first consider \eqref{eq:C112bbs}, which we will denote by
$C_{22}^{1,\mathrm{bbs}}$.  The corresponding contribution to the
spectral function can be cast in the form
\begin{equation}
\label{eq:bbs1}
S_{22}^{1,\mathrm{bbs}}(\omega,R,r)=
\frac{2\sigma_0^2\,\cot v\,\tan\frac{v}{2}}{\sqrt{(\omega-M\sin v)^2-M^2}}\,
\frac{\omega-2M\sin v}{\omega}\,
\cos\bigl(r\sqrt{(\omega-M\sin v)^2-M^2}\bigr)\,e^{2MR\cos v},
\end{equation}
and is seen to be exponentially suppressed in $MR$. We recall that the
parameter $v$ is defined as $\kappa=1-h^2/2M=\cos v$. In the limit
$v\rightarrow \pi/2$ the suppression of $S_{22}^{1,\mathrm{bbs}}$ becomes less
and less effective. In this limit the boundary bound state becomes weakly
bound and its effective size diverges~\cite{GhoshalZamolodchikov94}.

Similarly we obtain
\begin{equation}
\label{eq:bbs2}
\begin{split}
&S_{24}^{0,\mathrm{b},\mathrm{bbs}}(\omega,R,r)=\,
-\frac{2M^2\sigma_0^2\,\cot v\,\tan\frac{v}{2}}
{\sqrt{(\omega-M\sin v)^2-M^2}}\,
\frac{e^{2MR\cos v}}{\omega^2}\\*[2mm]
&\hspace{10mm}\cdot
\Biggl\{\biggl[
\frac{1}{2}\,\mathfrak{Re}K\bigl(\hat{\xi}+\ii\tfrac{\pi}{2}\bigr)\,
\bigl(\cosh 2\hat{\xi}-\cos 2v-2\bigr)-
2\,\mathfrak{Im}K\bigl(\hat{\xi}+\ii\tfrac{\pi}{2}\bigr)\,
\sinh\hat{\xi}\,\cos v\biggr]\cos\bigl(2MR\sinh\hat{\xi}\bigr)\\*[2mm]
&\hspace{11mm}
-\biggl[2\,\mathfrak{Re}K\bigl(\hat{\xi}+\ii\tfrac{\pi}{2}\bigr)\,
\sinh\hat{\xi}\,\cos v+
\frac{1}{2}\,\mathfrak{Im}K\bigl(\hat{\xi}+\ii\tfrac{\pi}{2}\bigr)\,
\bigl(\cosh 2\hat{\xi}-\cos 2v-2\bigr)\biggr]
\sin\bigl(2MR\sinh\hat{\xi}\bigr)\Biggr\},
\end{split}
\end{equation}
where
\begin{equation}
\label{eq:xihatbbs}
\hat{\xi}=\hat{\xi}(\omega,v)=
\text{Arcosh}\left(\frac{\omega}{M}-\sin v\right).  
\end{equation}
The contribution $S_{24}^{0,\mathrm{b},\mathrm{bbs}}$ is independent of $r$
and oscillates in $R$ with the frequency-dependent wave number
$2\sqrt{(\omega-M\sin v)^2-M^2}$.  In particular, for large $\omega$ this wave
number approaches $2(\omega-M\sin v)$. Furthermore, \eqref{eq:bbs2} is
exponentially suppressed for $R\rightarrow-\infty$. 

Interestingly, although both \eqref{eq:bbs1} and \eqref{eq:bbs2} have
threshold singularities at $\omega=M+M\sin v$, these singularities cancel each
other. This shows that we have to interpret their sum as the first-order
contribution of the boundary bound state to the spectral function, \ie we have
\begin{equation}
\label{eq:bbs3}
S_{1K}^\mathrm{bbs}(\omega,R,r)=
S_{22}^{1,\mathrm{bbs}}(\omega,R,r)+
S_{24}^{0,\mathrm{b},\mathrm{bbs}}(\omega,R,r).
\end{equation}
This indicates as well that the similar sum \eqref{eq:osc1K} should be viewed
as the oscillating contribution to first order in $K$. We observe that
$S_{1K}^\mathrm{bbs}$ has a gap $M+M\sin v$ which equals the energy of the
boundary bound state plus one additional particle.  On the other hand the
spectral function vanishes at $\omega=M\sin v$.  Combining these two
observations we conclude that
\begin{equation}
\bra{1_\mathrm{B}}\sigma^\mathrm{z}\ket{0_\mathrm{B}}=0 
\quad\text{but}\quad
\bra{1_\mathrm{B},\theta}\sigma^\mathrm{z}\ket{0_\mathrm{B}}\neq 0,
\end{equation}
where we have used the notations in the original picture with the boundary
located in space. The vanishing of the spectral function for
$\omega\rightarrow M+M\sin v$ means that the rapidity-dependence of the matrix
elements of $\sigma^\mathrm{z}$ overcomes the van-Hove type singularities in
the one-particle density of states at $\theta=0$.  

The result for free boundary conditions is obtained in the limit $v\rightarrow
0$. In this case \eqref{eq:bbs3} simplifies to
\begin{equation}
S_{1K,\mathrm{free}}^\mathrm{bbs}(\omega,R,r)=
\frac{\sigma_0^2\,e^{2MR}}{\sqrt{\omega^2-M^2}}
\left[\cos\bigl(r\sqrt{\omega^2-M^2}\bigr)
-\frac{M}{\omega}\left(\cos\bigl(2R\sqrt{\omega^2-M^2}\bigr)
+\frac{\sqrt{\omega^2-M^2}}{M}\sin\bigl(2R\sqrt{\omega^2-M^2}\bigr)
\right)\right]
\end{equation}
and the spectral gap equals $M$.

The second-order contribution in $K$ of the boundary bound state to the
spectral function is given by
\begin{equation}
\label{eq:bbs4}
S_{2K}^\mathrm{bbs}(\omega,R,r)=
S_{24}^{1,\mathrm{bbs}}(\omega,R,r)+S_{44}^{1,\mathrm{bbs}}(\omega,R,r)+
S_{26}^{0,\mathrm{bbs}}(\omega,R,r)+S_{46}^{0,\mathrm{bbs}}(\omega,R,r).
\end{equation}
Explicit expressions for the various contributions are given in
Appendix~\ref{sec:appbbs}. We find that all integrands appearing in
$S_{2K}^\mathrm{bbs}$ are exponentially small for large $M|R|$.  Furthermore,
$S_{2K}^\mathrm{bbs}$ remains finite at the lower threshold $\omega\rightarrow
M+M\sin v$, which is ensured by the cancellation of all singularities in
\eqref{eq:bbs4}. We expect similar cancellations of singularities to occur at
all orders. It is also straightforward to calculate the first correction to
\eqref{eq:bbs3} with more internal lines, which is given by
$C_{42}^{3,\mathrm{bbs}}+C_{44}^{2,\mathrm{b},\mathrm{bbs}}$. We find this to
be negligible for all values of $R$ and energies $\omega\le 25M$.

\subsection{Results for the local spectral function}
In the preceeding sections we have discussed individual terms in the expansion
of the spectral function. We are now in a position to combine the various
contributions and present results for the local spectral function of Ising
spins up to second order in the boundary $K$-matrix. If we truncate the
expansion at the two-particle level as well, the local spectral function
truncated is given by
\begin{eqnarray}
\label{eq:lsf}
S^\mathrm{lsf}(\omega,R)&=&S^\mathrm{lsf}_\mathrm{bulk}(\omega,R)+
S^\mathrm{lsf}_{1K}(\omega,R)+S^\mathrm{lsf}_{2K}(\omega,R)+\ldots\\*
S^\mathrm{lsf}_\mathrm{bulk}(\omega,R)&=&S^2_\mathrm{bulk}(\omega,R,r=0)
+\ldots\label{eq:lsfbulk}\\*
S^\mathrm{lsf}_{1K}(\omega,R)&=&S_{1K}^2(\omega,R,r=0)+
S_{1K}^\mathrm{osc}(\omega,R,r=0)+S_{1K}^\mathrm{bbs}(\omega,R,r=0)+\ldots
\label{eq:lsf1K}\\*
S^\mathrm{lsf}_{2K}(\omega,R)&=&S_{2K}^2(\omega,R,r=0)+
S_{2K}^\mathrm{osc}(\omega,R,r=0)+S_{2K}^\mathrm{bbs}(\omega,R,r=0)+\ldots
\label{eq:lsf2K}
\end{eqnarray}
where the terms originating in the boundary bound state are only present if
$h<h_\mathrm{c}$.  The dots in \eqref{eq:lsf} represent higher orders in $K$,
whereas the dots in \eqref{eq:lsfbulk}--\eqref{eq:lsf2K} stand for terms with
more than two particles in the intermediate state. As we have shown above, the
truncation at second order in $K$ is sufficient for $MR\alt -0.2$. We have
further argued that the truncation at the two-particle level gives a
very good accuracy for energies $\omega\le 25M$.

\begin{figure}[t]
\begin{center}
\includegraphics[scale=0.3,clip=true]{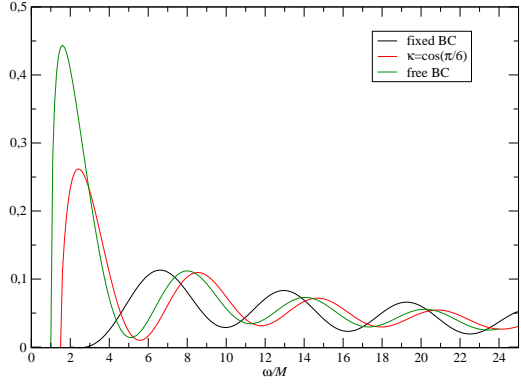}
\end{center}
\caption{Local spectral function \eqref{eq:lsf} for $MR=-0.5$ and fixed and 
  free boundary conditions as well as a boundary magnetic field corresponding
  to $v=\pi/6$ (in units of $\sigma_0^2/M$). The spectral function possesses a
  gap of $2M$ for fixed boundary conditions. For boundary magnetic fields
  $h<h_\mathrm{c}$ the gap is smaller than $2M$ and goes to $M$ in the limit
  of free boundary conditions.}
\label{fig:total}
\end{figure}
We have plotted \eqref{eq:lsf} in Fig.~\ref{fig:total} for fixed and free
boundary conditions as well as a boundary magnetic field corresponding to
$\kappa=\cos(\pi/6)$. For fixed boundary conditions the local spectral
function has a gap $2M$. The same behaviour of is well-known from the
system in the bulk.  For sufficient small values of the boundary magnetic
field, however, the gap is given by $E_1+M$, where $E_1$ is the energy of the
boundary bound state. This shows that the matrix element of
$\sigma^\mathrm{z}$ between the vacuum state in the presence of the boundary
and states containing the boundary bound state and one additional particle is
non-zero. In the limit of free boundary conditions $E_1$ tends to zero
and the gap approaches the single-particle gap $M$. Furthermore, we observe
oscillating behaviour in $\omega$ for all three boundary conditions.

\begin{figure}[t]
\begin{center}
\includegraphics[scale=0.3,clip=true]{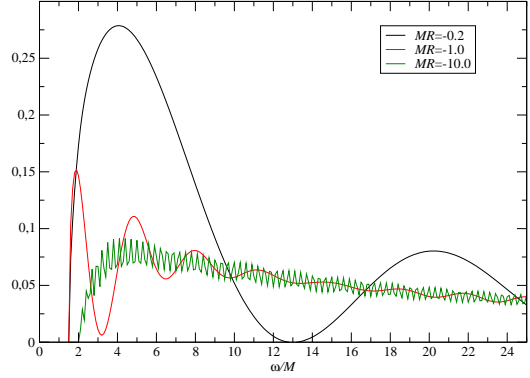}
\end{center}
\caption{Local spectral function \eqref{eq:lsf} for a boundary
  magnetic field corresponding to $v=\pi/6$ and different distances from the
  boundary (in units of $\sigma_0^2/M$).}
\label{fig:totalbbs}
\end{figure}
In Fig.~\ref{fig:totalbbs} we plot \eqref{eq:lsf} for a boundary
magnetic field corresponding to $\kappa=\cos(\pi/6)$. Close to the boundary the
spectral weight is concentrated at low energies and in particular at energies
$\omega\le 2M$. The weight below $2M$ is due to the creation of the boundary
bound state and one additional particle. The term responsible for this
behaviour is \eqref{eq:bbs3}.  Both \eqref{eq:bbs1} and \eqref{eq:bbs2} are
exponentially suppressed with increasing distance from the boundary. Hence the
spectral weight in the region $0\le \omega\le 2M$ is very small, as
can be seen in Fig.~\ref{fig:totalbbs} for $MR=-10$. This strong suppression
may complicate an experimental detection of this effect of the boundary bound
state on the local spectral function. Furthermore, the spectral function
oscillates in $\omega$ with a frequency depending on the distance from the
boundary. For large $|R|$ these oscillations become very rapid, which may
cause difficulties in detecting this effect as well.

\begin{figure}[t]
\begin{center}
\includegraphics[scale=0.3,clip=true]{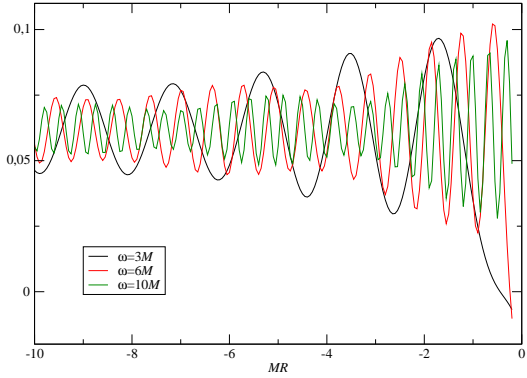}
\end{center}
\caption{Local spectral function \eqref{eq:lsf} for fixed boundary conditions
  as function of the distance from the boundary (in units of $\sigma_0^2/M$).}
\label{fig:totalfixedw}
\end{figure}
Finally, we plot the local spectral function as a function of the
distance from the boundary in Fig.~\ref{fig:totalfixedw}. It is clear from this
plot that the oscillations are only algebraically decaying in $|R|$. 

In summary, the local spectral function \eqref{eq:lsf} exhibits two
new features in the presence of a boundary. First, the spectral
function shows oscillatory behaviour both as a function of $\omega$
with a frequency depending on the distance from the boundary, and
as a function of $R$ with an energy dependent frequency. Second, for
values of the boundary magnetic field below the 
critical value $h_\mathrm{c}=\sqrt{2M}$ we have found spectral weight in the
interval $M\le\omega\le 2M$, \ie within the gap of the bulk Ising
model. This spectral weight is due to the existence of a boundary bound state,
and hence strongly suppressed with increasing distance from the boundary.

Any quasi one-dimensional material described by the quantum Ising
model needs to be thought of as an ensemble of finite length chains.
If the average chain length is large, a model in terms of
semi-infinite chains should constitute a good starting point and the
results obtained here should be applicable. In particular,
it should be possible to detect the midgap states in inelastic neutron
scattering experiments. On the other hand, the oscillatory behaviour
of dynamical spin correlations is likely to average out.

\section{Correlation function of the disorder field}
\label{sec:disordercorr}
In this section we briefly discuss the two-point function of the disorder
field $\mu^\mathrm{z}$. Up to first order in $K$ we find
\begin{eqnarray}
\bra{0_\mathrm{B}}\mathcal{T}_\tau\,\mu^\mathrm{z}(\tau,x_1)\,
\mu^\mathrm{z}(0,x_2)\ket{0_\mathrm{B}}\!\!&=&\!\!
\bra{0}\mathcal{T}_x\,\mu^\mathrm{z}(\tau,x_1)\,
\mu^\mathrm{z}(0,x_2)\ket{\mathrm{B}}\nonumber\\*
&=&\!\!
\bra{0}\mu^\mathrm{z}(\tau,x_1)\,\mu^\mathrm{z}(0,x_2)\ket{0}
+\int_0^\infty\frac{d\xi}{2\pi}\,K(\xi)\,
\bra{0}\mu^\mathrm{z}(\tau,x_1)\,\mu^\mathrm{z}(0,x_2)\ket{-\xi,\xi}+\ldots,
\label{eq:mucorr}
\end{eqnarray}
where we have assumed that $x_1<x_2$. The first term equals the result
in the bulk. The single-particle contribution to it is
\begin{equation}
\label{eq:mu0}
\sigma_0^2\int_{-\infty}^\infty\frac{d\theta}{2\pi}\,e^{-Mr\cosh\theta}\,
e^{\ii M\tau\sinh\theta}=
\sigma_0^2\int_{-\infty}^\infty\frac{d\theta}{2\pi}\,
e^{-M\sqrt{r^2+\tau^2}\cosh\theta}=
\frac{\sigma_0^2}{\pi}\,K_0\bigl(M\sqrt{r^2+\tau^2}\bigr),
\end{equation}
where we have used \eqref{eq:appcontinuation}.  After analytic continuation,
$\tau=\ii t$, we observe typical light-cone behaviour. Fourier transforming we
find that the single-particle contribution to the bulk part of the spectral
function is given by
\begin{equation}
\label{eq:Smubulk}
S_\mathrm{bulk}^{\mu^\mathrm{z}}(\omega>M,r>0)=
\frac{\sigma_0^2}{\sqrt{\omega^2-M^2}}\,\cos\Bigl(r\sqrt{\omega^2-M^2}\Bigr).
\end{equation}
We note that the spectral function shows undamped oscillations as function of 
$r$.

If we assume $\tau>0$, the leading contribution due to the presence of the
boundary can be calculated using \eqref{eq:reg1}
\begin{equation}
\bra{\theta}\mu^\mathrm{z}\ket{-\xi,\xi}=
\bra{\theta\pm\ii 0}\mu^\mathrm{z}\ket{-\xi,\xi}+
2\pi\sigma_0\,\delta(\theta-\xi)-2\pi\sigma_0\,\delta(\theta+\xi).
\end{equation}
The connected piece merely yields a correction to the one-particle spectral
function \eqref{eq:Smubulk} which vanishes for large distances from the
boundary. However, the disconnected pieces give
\begin{eqnarray}
\sigma_0^2\int_{-\infty}^\infty\frac{d\xi}{2\pi}\,K(\xi)\,
e^{2MR\cosh\theta}\,e^{\ii M\tau\sinh\theta}&=&
\sigma_0^2\int_{-\infty}^\infty\frac{d\xi}{2\pi}\,K(\xi+\ii\xi_0)\,
e^{-M\sqrt{4R^2+\tau^2}\cosh\xi}\label{eq:muf1}\\*[2mm]
& &+\Theta(h_{\rm c}-h)\ \Theta(\xi_0-v)\ 
2\sigma_0^2\,\cot v\,\tan\frac{v}{2}\,e^{2MR\cos v}\,e^{-M\tau\sin v},
\label{eq:muf2}
\end{eqnarray}
where $\xi_0=\arctan(\tau/2|R|)$. Note that \eqref{eq:muf1} and
\eqref{eq:muf2} are independent of $r$.  If we perform the analytic
continuation $\tau=\ii t$ in \eqref{eq:muf1}, we find an exponentially damped
behaviour for $4R^2>t^2$, and an oscillating behaviour for $4R^2<t^2$.  The
contribution to the spectral function corresponding to \eqref{eq:muf1} is
given by
\begin{equation}
\label{eq:Smuf1}
S_\mathrm{1K}^{\mu^\mathrm{z},\mathrm{osc}}
=\frac{\sigma_0^2}{\sqrt{\omega^2-M^2}}\,
\Bigl[
\mathfrak{Re}\,K\bigl(\text{Arcosh}\tfrac{\omega}{M}+\ii\tfrac{\pi}{2}\bigr)\,
\cos\bigl(2\sqrt{\omega^2-M^2}\,R\bigr)-
\mathfrak{Im}\,K\bigl(\text{Arcosh}\tfrac{\omega}{M}+\ii\tfrac{\pi}{2}\bigr)\,
\sin\bigl(2\sqrt{\omega^2-M^2}\,R\bigr)\Bigr].
\end{equation}
We observe that \eqref{eq:Smuf1} is not damped for large distances $|R|$. This
is due to the dissipationless propagation of a single particle to the boundary
and back, and a likewise dissipationless reflection off the boundary. On the
other hand, \eqref{eq:muf2} represents the contribution of the boundary bound
state.  It is seen to give rise to a $\delta$-peak contribution to the
spectral function at the energy $\omega=M\sin v$, which implies that
$\bra{1_\mathrm{B}}\mu^\mathrm{z}\ket{0_\mathrm{B}}\neq 0$. Higher order
contributions to the two-point function of disorder operators can be
calculated by the same method we employed above for the spin correlations.

\section{Conclusions}
Our main result is the calculation of the dynamical spin-spin
correlations in the Ising field theory with a boundary. We have
derived an expansion in powers of the boundary reflection matrix $K$
and shown that it converges rapidly even in the case where both
operators in the two-point function are fairly close to the boundary. 
We have also demonstrated that like in the bulk case higher-order terms in
the expansion in the number of particles in the intermediate state are
negligible at low energies. The most notable effect of the boundary is
that at sufficiently late times the spin-spin correlations show
oscillatory behaviour arbitrarily far away from the boundary. As is
well known, for small values of the boundary magnetic field a boundary
bound state exists. This bound state leads to a contribution to the
spectral function within the gap of the bulk Ising chain. Similar 
features are also found for the Green's functions of the Majorana
fermions and the two-point function of the disorder field. 

We have seen that the expansion in powers of the boundary $K$-matrix
breaks down close to the boundary. In order to access this regime,
other methods are necessary. One possible approach is the truncated
conformal space approach, which has already been applied successfully
to the analogous problem for one-point functions~\cite{Dorey-01}. In
would be interesting to generalise these results to the case of
two-point functions and in this way obtain accurate expressions for
the two-point function for all values of $MR$.

The results obtained in the present work have applications not only to
the quantum Ising model itself. It is well-known that both anisotropic
spin-1 Heisenberg chains~\cite{AMT} and the weak-coupling two-leg
spin-1/2 ladder~\cite{SNT} can be described in terms of three and four
Ising models respectively. 
Correlation functions of the staggered components of the spin
operators in these models are represented by products of two-point
functions of the spin and disorder fields of the quantum Ising chain.
Using the results obtained in the present work one may calculate
dynamical correlation functions for these systems.

Finally, having established that the expansion in powers of the
boundary $K$-matrix converges even relatively close to the boundary, one
may apply the method used here to other systems. In a forthcoming work
we apply the method used in the present work to the calculation of the
local tunneling density of states in a one-dimensional charge density
wave state~\cite{unpub}. 

\section*{Acknowledgements}
We would like to thank Joe Bhaseen, Eduardo Fradkin, Robert Konik and
Philippe Lecheminant for useful discussions. DS was supported by the
Deutsche Akademie der Naturforscher Leopoldina by means of the
Bundesministerium f\"ur Bildung und Forschung under grant no BMBF-LPD
9901/8-145. FHLE acknowledges support by the EPSRC under grant
EP/D050952/1. 

\appendix
\section{Useful formulas}\label{sec:appuf}
We summarise some formulas, which were frequently used in the derivation of
the $C_{2n\,2m}$'s.
\begin{eqnarray}
\sinh(x\pm\ii\pi/2)&=&\pm\ii\cosh x,\\[2mm]
\cosh(x\pm\ii\pi/2)&=&\pm\ii\sinh x,\\[2mm]
\tanh\frac{x-y}{2}\,\tanh\frac{x+y}{2}&=&
\frac{\cosh x-\cosh y}{\cosh x+\cosh y},\\[2mm]
\tanh\frac{x-y\mp\ii\pi/2}{2}\,\tanh\frac{x+y\pm\ii\pi/2}{2}&=&
\frac{\cosh x\mp\ii\sinh y}{\cosh x\pm\ii\sinh y}\\[2mm]
\tanh\frac{x-y\mp\ii\pi/2}{2}\,\coth\frac{x+y\pm\ii\pi/2}{2}&=&
-\frac{\cosh y\pm\ii\sinh x}{\cosh y\mp\ii\sinh x},\\[2mm]
\coth\frac{x-y\mp\ii\pi/2}{2}\,\coth\frac{x+y\pm\ii\pi/2}{2}&=&
\frac{\cosh x\pm\ii\sinh y}{\cosh x\mp\ii\sinh y}.
\end{eqnarray}

If $x,y\in\mathbb{R}$, $y>0$, $f(\theta)$ is analytic in
$-\arctan(x/y)\le\mathfrak{Im}\,\theta\le\arctan(x/y)$ and at most
exponentially growing for $\theta\rightarrow\pm\infty$, then
\begin{equation}
\int_{-\infty}^\infty\frac{d\theta}{2\pi}\,f(\theta)\,e^{\ii Mx\sinh\theta}\,
e^{-My\cosh\theta}=
\int_{-\infty}^\infty\frac{d\theta}{2\pi}\,f(\theta+\ii\theta_0)\,
e^{-M\sqrt{x^2+y^2}\cosh\theta},
\label{eq:appcontinuation}
\end{equation}
where $\theta_0=\arctan(x/y)$. In particular, if $\alpha\in\mathrm{R}$
and $f(\theta)=e^{\alpha\theta}$ then.  
\begin{equation}
\int_{-\infty}^\infty\frac{d\theta}{2\pi}\,e^{\alpha\theta}\,
e^{\ii Mx\sinh\theta}\,e^{-My\cosh\theta}=
\left(\frac{\ii y-x}{\ii y+x}\right)^{\alpha/2}
\int_{-\infty}^\infty\frac{d\theta}{2\pi}\,e^{\alpha\theta}\,
e^{-M\sqrt{x^2+y^2}\cosh\theta}.
\label{eq:appcontinuationspecial}
\end{equation}
If $f$ possesses poles in the strip
$-\arctan(x/y)\le\mathfrak{Im}\,\theta\le\arctan(x/y)$ additional
contributions appear on the right-hand side of \eqref{eq:appcontinuation}.

The explicit form of the boundary reflection matrix $K$ at
$\xi+\ii\pi/2$ is
\begin{equation}
\label{eq:defKhpi2}
\begin{split}
K\bigl(\xi+\ii\tfrac{\pi}{2}\bigr)=&\frac{1}{1+\tanh^2\xi/2}
\frac{1}{\kappa^2+\sinh^2\xi}\left[
\left(\tanh^2\frac{\xi}{2}-1\right)\bigl(\kappa^2-\sinh^2\xi\bigr)
-4\kappa\tanh\frac{\xi}{2}\,\sinh\xi\right.\\[2mm]
&\hspace{30mm}\left.
+2\ii\left(\tanh\frac{\xi}{2}\,\bigl(\kappa^2-\sinh^2\xi\bigr)
+\kappa\sinh\xi\left(\tanh^2\frac{\xi}{2}-1\right)
\right)
\right].
\end{split}
\end{equation}
For the special cases of free and fixed boundary conditions this
simplifies to
\begin{eqnarray}
\label{eq:defKfree}
K_\mathrm{free}\bigl(\xi+\ii\tfrac{\pi}{2}\bigr)&=&
\frac{1}{1+\coth^2\xi/2}\left[
1-\coth^2\frac{\xi}{2}-2\ii\coth\frac{\xi}{2}
\right],\label{eq:defKfreepi2}\\[2mm]
K_\mathrm{fixed}\bigl(\xi+\ii\tfrac{\pi}{2}\bigr)&=&
-\frac{1}{1+\tanh^2\xi/2}\left[
1-\tanh^2\frac{\xi}{2}-2\ii\tanh\frac{\xi}{2}
\right].\label{eq:defKfixedpi2}
\end{eqnarray}
We note that the real part is even under $\xi\rightarrow-\xi$ whereas the
imaginary part is odd. 

In the calculation of the spectral functions we use that if
$\omega\in\mathbb{R}$, $f(\theta_1,\theta_2)=-f(-\theta_1,-\theta_2)$ and $f$
sufficiently well behaved, then
\begin{eqnarray}
\int_{-\infty}^\infty d\theta_1\,
\mathcal{P}\!\!\!\int_{-\infty}^\infty d\theta_2\,
\frac{f(\theta_1,\theta_2)}{\omega-\sum_i\cosh\theta_i}&=&0,\\[2mm]
\int_{-\infty}^\infty d\theta_1 d\theta_2\,f(\theta_1,\theta_2)
\,\delta\Bigl(\omega-\sum\nolimits_i\cosh\theta_i\Bigr)&=&0.
\end{eqnarray}
Furthermore, we use
\begin{eqnarray}
\mathfrak{Re}\,
\prod_{i=1}^2\frac{\cosh\xi+\ii\sinh\theta_i}{\cosh\xi-\ii\sinh\theta_i}&=&
\frac{\prod_i^2(\cosh^2\xi-\sinh^2\theta_i)
-4\cosh^2\xi\,\prod_i^2\sinh\theta_i}
{\prod_i^2(\cosh^2\xi+\sinh^2\theta_i)},\label{eq:appRe}\\*[2mm]
\mathfrak{Im}\,
\prod_{i=1}^2\frac{\cosh\xi+\ii\sinh\theta_i}{\cosh\xi-\ii\sinh\theta_i}&=&
2\cosh\xi\,
\frac{\sum_i^2\sinh\theta_i\prod_{j\neq i}^2(\cosh^2\xi-\sinh^2\theta_j)}
{\prod_i^2(\cosh^2\xi+\sinh^2\theta_i)},\label{eq:appIm}
\end{eqnarray}
and similar formulas for the higher-order terms.  Note that \eqref{eq:appRe}
is even under simultaneous reflection $\theta_i\rightarrow-\theta_i$,
while \eqref{eq:appIm} is odd.

\section{Finite-size regularisation of $\bs{C_{22}}$}\label{sec:finite}
In order to check the infinite volume regularisation scheme for
form factors involving two multiparticle states, we
evaluate $C_{22}$ in the finite system $0\le\tau\le L$. A similar
analysis was performed recently for the correlation functions in the
Ising model at finite temperatures~\cite{EsslerKonik07}.  The form
factors for $\sigma^\mathrm{z}$ in the finite system
are~\cite{FonsecaZamolodchikov03} 
\begin{equation}
\label{eq:finiteff}
_\mathrm{NS}\!\bra{k_1,\ldots,k_m}\sigma^\mathrm{z}
\ket{n_1,\ldots,n_n}_\mathrm{R}=
S(L)\prod_{i=1}^m\tilde{g}(\xi_{k_i})\prod_{j=1}^ng(\theta_{n_j})
F_{m,n}(\xi_{k_1},\ldots,\xi_{k_m}|\theta_{n_1},\ldots,\theta_{n_n}),
\end{equation}
where the rapidities in the Neveu--Schwarz (NS) and Ramond (R) sector are
given by
\begin{eqnarray}
\text{NS}:& &ML\sinh\xi_k=2\pi k,\; k\in\mathbb{Z}+\tfrac{1}{2},\\
\text{R}:& &ML\sinh\theta_n=2\pi n,\; n\in\mathbb{Z}.
\end{eqnarray}
The function $F_{m,n}$ in \eqref{eq:finiteff} is the infinite-volume form
factor
\begin{equation}
F_{m,n}(\xi_{1},\ldots,\xi_{m}|\theta_{1},\ldots,\theta_{n})=
\ii^{\lfloor(m+n)/2\rfloor}\sigma_0 
\prod_{\substack{i,j=1\\ i<j}}^m\tanh\frac{\xi_i-\xi_j}{2}
\prod_{\substack{i,j=1\\ i<j}}^n\tanh\frac{\theta_i-\theta_j}{2}
\prod_{i=1}^m\prod_{j=1}^n\coth\frac{\xi_i-\theta_j}{2},
\label{eq:finiteF}
\end{equation}
where the symbol $\lfloor \; \rfloor$ denotes the floor function, \ie $\lfloor
x\rfloor$ is the largest integer $l\le x$, and the constant as well as the leg
factors are
\begin{equation}
S(L)=1+\mathcal{O}(e^{-L}),\quad
g(\theta)=\tilde{g}(\theta)=\frac{1}{\sqrt{ML\cosh\theta}}+\mathcal{O}(e^{-L}).
\end{equation}
We note that due to the fact that $\sigma^\mathrm{z}$ connects the NS and R
sectors of the Hilbert space, the rapidities $\xi_i$ and $\theta_j$ cannot
coincide and therefore no singularities occur in the finite volume.

In terms of these form factors the finite volume regularisation of
$C_{22}$ is given by (we assume $x_1<x_2$ for simplicity)
\begin{eqnarray}
\label{eq:finiteC}
C_{22}&=&
\frac{1}{2}\sum_{\substack{k\in\mathrm{NS}\\ k>0}}\;
\sum_{n_1,n_2\in\mathrm{R}} K(\xi_k)\,
_\mathrm{NS}\!\bra{0}\sigma^\mathrm{z}(\tau,x_1)
\ket{n_1,n_2}_\mathrm{R}\!\!\!\!\!\!
\phantom{\ket{0}}_\mathrm{R}\!\bra{n_2,n_1}\sigma^\mathrm{z}(0,x_2)
\ket{-k,k}_\mathrm{NS}\label{eq:finiteC2}\\[2mm]
&=&-\ii\frac{\sigma_0^2}{2} S(L)^2
\sum_{\substack{k\in\mathrm{NS}\\ k>0}}\;\sum_{n_1,n_2\in\mathrm{R}} 
K(\xi_k)\,\,e^{2Mx_2\cosh\xi_k}\,
e^{-Mr\sum_i\cosh\theta_{n_i}}\,e^{\ii M\tau\sum_i\sinh\theta_{n_i}}
\nonumber\\*
& &\hspace{15mm}\cdot
\tilde{g}^2(\xi_k)\,\tanh\xi_k\,
\tanh^2\frac{\theta_{n_1}-\theta_{n_2}}{2}\,
\prod_{i=1}^2 g^2(\theta_{n_i})\,
\coth\frac{\xi_k-\theta_{n_i}}{2}\coth\frac{\xi_k+\theta_{n_i}}{2}.
\label{eq:C11finite}
\end{eqnarray}
We have evaluated \eqref{eq:C11finite} and \eqref{eq:C_11} numerically for
several values of $x_1$, $x_2$ and $0.2\le \tau\le 1$ as well as different
values of the boundary magnetic field. The two expressions coincide within a
relative error of less than $10^{-4}$. For the evaluation we used $L=50$ and
$N=350$ as cut-off for the momenta, \ie $0<k<N$ and $-N\le n_i\le N$. The
result does not change if we use Ramond states as the outer states in
\eqref{eq:finiteC} instead of Neveu--Schwarz ones.

\section{Higher-order corrections to (\ref{eq:corr})}\label{sec:appcorr}
In this appendix we calculate several higher-order terms in the expansion
\eqref{eq:corrmn}, which can be compared with the leading contributions
calculated in Sec.~\ref{sec:corr}. In order to simplify the notations, we will
restrict ourselves to $\tau>0$ and $r>0$. The correlation function for general
$\tau$ and $r$ can be obtained as in Sec.~\ref{sec:corr}.

\subsection{Calculation of $\bs{C_{42}}$}
The next term in the series \eqref{eq:corr} is given by
\begin{equation}
\begin{split}
C_{42}(\tau,x_1,x_2)=&\;\frac{1}{4!}\int_0^\infty\frac{d\xi}{2\pi}
\int_{-\infty}^\infty\frac{d\theta_1 d\theta_2 d\theta_3 d\theta_4}{(2\pi)^4}
K(\xi)\,e^{2M\max(x_1,x_2)\cosh\xi}e^{-M|r|\sum_i\cosh\theta_i}\\[2mm]
&\hspace{20mm}\cdot e^{\ii \,\sgn{r}M\tau\sum_i\sinh\theta_i}
f(\theta_1,\theta_2,\theta_3,\theta_4)
\bra{\theta_4,\theta_3,\theta_2,\theta_1}\sigma^\mathrm{z}\ket{-\xi,\xi}.
\label{eq:C21}
\end{split}
\end{equation}
\begin{figure}[t]
\begin{center}
\includegraphics[scale=0.3,clip=true]{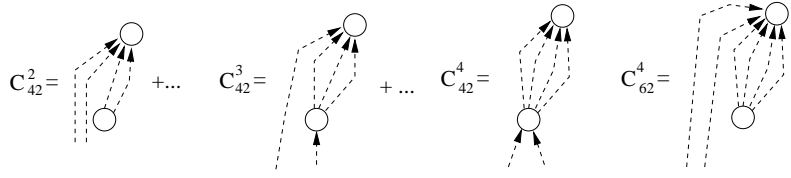}
\end{center}
\caption{Graphical representation of $C_{42}$ and $C_{62}^4$. The dots 
represent diagrams with the same number of contractions.}
\label{fig:diagrams3}
\end{figure}
As we restrict ourselves to $\tau>0$ and $r>0$, we can apply the
regularisation scheme \eqref{eq:reg1} to evaluate the second matrix element
\begin{equation}
\begin{split}
&\hspace{-10mm}
\bra{\theta_4,\theta_3,\theta_2,\theta_1}\sigma^\mathrm{z}\ket{-\xi,\xi}=
\bra{\theta_4,\theta_3}-\xi,\xi\rangle\,
f(\theta_2+\ii \pi+\ii\eta_2,\theta_1+\ii \pi+\ii\eta_1)\\*
&-\bra{\theta_4,\theta_2}-\xi,\xi\rangle\,
f(\theta_3+\ii \pi+\ii\eta_3,\theta_1+\ii \pi+\ii\eta_1)\\*
&+\bra{\theta_4,\theta_1}-\xi,\xi\rangle\,
f(\theta_3+\ii \pi+\ii\eta_3,\theta_2+\ii \pi+\ii\eta_2)\\*
&+\bra{\theta_3,\theta_2}-\xi,\xi\rangle\,
f(\theta_4+\ii \pi+\ii\eta_4,\theta_1+\ii \pi+\ii\eta_1)\\*
&-\bra{\theta_3,\theta_1}-\xi,\xi\rangle\,
f(\theta_4+\ii \pi+\ii\eta_4,\theta_2+\ii \pi+\ii\eta_2)\\*
&+\bra{\theta_2,\theta_1}-\xi,\xi\rangle\,
f(\theta_4+\ii \pi+\ii\eta_4,\theta_3+\ii \pi+\ii\eta_3)\\
&-\bra{\theta_4}\xi\rangle\,
f(\theta_3+\ii \pi+\ii \eta_3,\theta_2+\ii \pi+\ii \eta_2,
\theta_1+\ii \pi+\ii \eta_1,-\xi)\\*
&+\bra{\theta_3}\xi\rangle\,
f(\theta_4+\ii \pi+\ii \eta_4,\theta_2+\ii \pi+\ii \eta_2,
\theta_1+\ii \pi+\ii \eta_1,-\xi)\\*
&-\bra{\theta_2}\xi\rangle\,
f(\theta_4+\ii \pi+\ii \eta_4,\theta_3+\ii \pi+\ii \eta_3,
\theta_1+\ii \pi+\ii \eta_1,-\xi)\\*
&+\bra{\theta_1}\xi\rangle\,
f(\theta_4+\ii \pi+\ii \eta_4,\theta_3+\ii \pi+\ii \eta_3,
\theta_2+\ii \pi+\ii \eta_2,-\xi)\\
&+\bra{\theta_4}-\xi\rangle\,
f(\theta_3+\ii \pi+\ii \eta_3,\theta_2+\ii \pi+\ii \eta_2,
\theta_1+\ii \pi+\ii \eta_1,\xi)\\*
&-\bra{\theta_3}-\xi\rangle\,
f(\theta_4+\ii \pi+\ii \eta_4,\theta_2+\ii \pi+\ii \eta_2,
\theta_1+\ii \pi+\ii \eta_1,\xi)\\*
&+\bra{\theta_2}-\xi\rangle\,
f(\theta_4+\ii \pi+\ii \eta_4,\theta_3+\ii \pi+\ii \eta_3,
\theta_1+\ii \pi+\ii \eta_1,\xi)\\*
&-\bra{\theta_1}-\xi\rangle\,
f(\theta_4+\ii \pi+\ii \eta_4,\theta_3+\ii \pi+\ii \eta_3,
\theta_2+\ii \pi+\ii \eta_2,\xi)\\
&+
f(\theta_4+\ii \pi+\ii \eta_4,\theta_3+\ii \pi+\ii \eta_3,
\theta_2+\ii \pi+\ii \eta_2,\theta_1+\ii \pi+\ii \eta_1,
-\xi,\xi),
\end{split}
\label{eq:regC21}
\end{equation}
This yields three terms, which are graphically represented in
Fig.~\ref{fig:diagrams3}. The first one is found to be
$C_{42}^2(\tau,x_1,x_2)=C_{22}^2(\tau,x_2,x_1)$ for general $x_1$ and $x_2$.
Furthermore, the contributions due to lines seven through 14 give
\begin{eqnarray}
C_{42}^3(\tau,x_1,x_2)&=&-\frac{\sigma_0^2}{6}\int_{-\infty}^\infty
\frac{d\xi}{2\pi}\frac{d\theta_1d\theta_2d\theta_3}{(2\pi)^3}\,
K(\xi)\,\prod_{i=1}^3\tanh\frac{\xi-\theta_i}{2}\,
\coth\frac{\xi+\theta_i+\ii\eta_i}{2}\nonumber\\*
& &\hspace{15mm}\cdot 
\prod_{\substack{i,j=1\\ i<j}}^{3}\tanh^2\frac{\theta_i-\theta_j}{2}\,
e^{2MR\cosh\xi}\,e^{-Mr\sum_i\cosh\theta_i}\,
e^{\ii M\tau(\sinh\xi+\sum_i\sinh\theta_i)}\\[2mm]
&=&-\frac{\sigma_0^2}{6}\int_{-\infty}^\infty
\frac{d\xi}{2\pi}\frac{d\theta_1d\theta_2d\theta_3}{(2\pi)^3}\,
K\bigl(\xi+\ii\tfrac{\pi}{2}\bigr)\,
\prod_{i=1}^3\frac{\cosh\xi-\cosh\theta_i}{\cosh\xi+\cosh\theta_i}
\nonumber\\*
& &\hspace{15mm}\cdot 
\prod_{\substack{i,j=1\\ i<j}}^{3}\tanh^2\frac{\theta_i-\theta_j}{2}\,
e^{\ii 2MR\sinh\xi}\,e^{-\ii Mr\sum_i\sinh\theta_i}\,
e^{-M\tau(\cosh\xi+\sum_i\cosh\theta_i)}
\label{eq:C212osc}\\*[2mm]
& &+\Theta(h_{\rm c}-h)\ 
\frac{\sigma_0^2}{3}\,\cot v\,\tan\frac{v}{2}\,e^{2MR\cos v}\,
e^{-M\tau\sin v}
\int_{-\infty}^\infty\frac{d\theta_1d\theta_2d\theta_3}{(2\pi)^3}\,
\prod_{i=1}^3\frac{\cosh\theta_i-\sin v}{\cosh\theta_i+\sin v}\,\nonumber\\*
& &\hspace{30mm}\cdot
\prod_{\substack{i,j=1\\ i<j}}^{3}\tanh^2\frac{\theta_i-\theta_j}{2}\,
e^{-\ii Mr\sum_i\sinh\theta_i}\,e^{-M\tau\sum_i\cosh\theta_i}.
\label{eq:C212bbs}
\end{eqnarray}
Here \eqref{eq:C212osc} shows the typical oscillating behaviour
and \eqref{eq:C212bbs} is present if $h<h_\mathrm{c}$.

Finally, the last term of \eqref{eq:regC21} yields
\begin{equation}
\begin{split}
C_{42}^4(\tau,x_1,x_2)
&\,=-\ii \frac{\sigma_0^2}{24}\int_0^\infty\frac{d\xi}{2\pi}
\int_{-\infty}^\infty\frac{d\theta_1d\theta_2d\theta_3d\theta_4}{(2\pi)^4} 
\,K(\xi)\,\tanh\xi\,
\prod_{\substack{i,j=1\\ i<j}}^{4}\tanh^2\frac{\theta_i-\theta_j}{2}\\*
&\hspace{25mm}\cdot \prod_{i=1}^4
\frac{\cosh\xi+\ii \sinh\theta_i}{\cosh\xi-\ii \sinh\theta_i}\,
e^{2Mx_2\cosh\xi}\,e^{-\ii Mr\sum_i\sinh\theta_i}\,
e^{-M\tau\sum_i\cosh\theta_i}.
\end{split}
\end{equation}
The next term in the expansion \eqref{eq:corr} possesses a similar leading
term, formally we find $C_{62}^4(\tau,x_1,x_2)=C_{42}^4(\tau,x_2,x_1)$.

\subsection{Calculation of $\bs{C_{24}}$}
The fourth to seventh line of \eqref{eq:regC12} yield
\begin{eqnarray}
C_{24}^1(\tau,x_1,x_2)&=&\ii \sigma_0^2
\int_0^\infty\frac{d\xi_1}{2\pi}
\int_{-\infty}^\infty\frac{d\xi_2}{2\pi}\frac{d\theta}{2\pi}\,
K(\xi_1)\,K\bigl(\xi_2+\ii\tfrac{\pi}{2}\bigr)\,\tanh\xi_1\nonumber\\*[2mm]
& &\hspace{15mm}\cdot
\frac{\cosh\xi_1-\ii\sinh\xi_2}{\cosh\xi_1+\ii\sinh\xi_2}\,
\frac{\cosh\xi_1+\ii \sinh\theta}{\cosh\xi_1-\ii \sinh\theta}\,
\frac{\cosh\xi_2-\cosh\theta}{\cosh\xi_2+\cosh\theta}\nonumber\\*[2mm]
& &\hspace{15mm}\cdot
e^{2Mx_2\cosh\xi_1}\,e^{\ii 2MR\sinh\xi_2}\,e^{-\ii Mr\sinh\theta}\,
e^{-M\tau(\cosh\theta+\cosh\xi_2)}\label{eq:C123osc}\\*[2mm]
& &-\Theta(h_{\rm c}-h)\ 
2\ii\sigma_0^2\,\cot v\,\tan\frac{v}{2}\,e^{2MR\cos v}\,
e^{-M\tau\sin v}\nonumber\\*
& &\hspace{10mm}\cdot
\int_0^\infty\frac{d\xi}{2\pi}\int_{-\infty}^\infty\frac{d\theta}{2\pi}\,
K(\xi)\,\tanh\xi\,\frac{\cosh\xi-\cos v}{\cosh\xi+\cos v}
\frac{\cosh\theta-\sin v}{\cosh\theta+\sin v}\nonumber\\*[2mm]
& &\hspace{15mm}\cdot
\frac{\cosh\xi+\ii \sinh\theta}{\cosh\xi-\ii \sinh\theta}\,
e^{2Mx_2\sinh\xi}\,e^{-\ii Mr\sinh\theta}\,
e^{-M\tau\cosh\theta}.\label{eq:C123bbs}
\end{eqnarray}
We stress that the integrand in \eqref{eq:C123osc} is exponentially suppressed
for $x_2\rightarrow-\infty$.  Furthermore, the last term in \eqref{eq:regC12}
yields
\begin{equation}
\begin{split}
C_{24}^2(\tau,x_1,x_2)=
&\,-\frac{\sigma_0^2}{4}\int_0^\infty\frac{d\xi_1d\xi_2}{(2\pi)^2}
\int_{-\infty}^\infty\frac{d\theta_1 d\theta_2}{(2\pi)^2}
\prod_{i=1}^2 K(\xi_i)\,\tanh\xi_i\,
\tanh^2\frac{\theta_1-\theta_2}{2}\,
\left(\frac{\cosh\xi_1-\cosh\xi_2}{\cosh\xi_1+\cosh\xi_2}\right)^2\\*
&\hspace{20mm}\cdot
\prod_{i,j=1}^{2}\frac{\cosh\xi_i+\ii \sinh\theta_j}
{\cosh\xi_i-\ii \sinh\theta_j}
e^{2Mx_2\sum_i\cosh\xi_i}\,
e^{-\ii Mr\sum_i\sinh\theta_i}\,
e^{-M\tau\sum_i\cosh\theta_i}.
\end{split}
\end{equation}

\subsection{Calculation of $\bs{C_{44}}$}
\begin{figure}[t]
\begin{center}
\includegraphics[scale=0.3,clip=true]{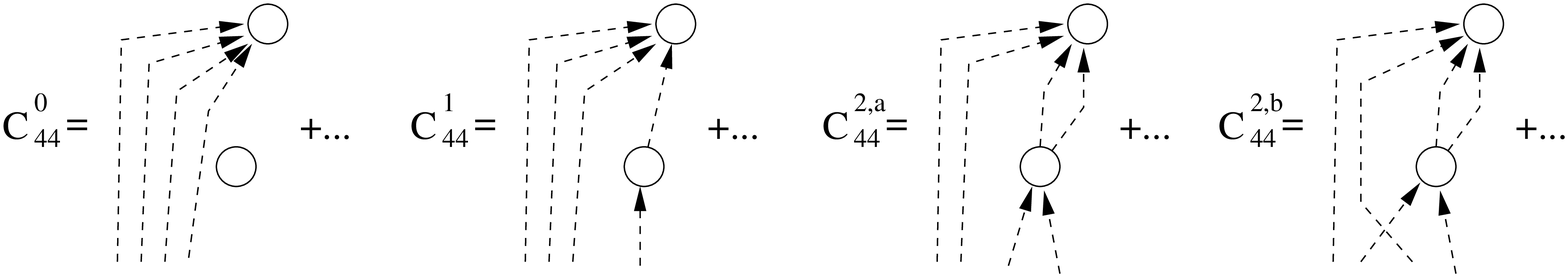}
\end{center}
\caption{Graphical representation of $C_{44}$.}
\label{fig:diagrams6}
\end{figure}
For the evaluation of $C_{44}$ we have 
\begin{equation}
\begin{split}
C_{44}(\tau,x_1,x_2)=&\;\frac{1}{48}\int_0^\infty\frac{d\xi_1d\xi_2}{(2\pi)^2}
\int_{-\infty}^\infty\frac{d\theta_1 d\theta_2 d\theta_3 d\theta_4}{(2\pi)^4}\,
K(\xi_1)\,K(\xi_2)\,e^{2Mx_2\sum_i\cosh\xi_i}\\[2mm]
&\hspace{20mm}\cdot 
e^{-Mr\sum_i\cosh\theta_i}\,e^{\ii M\tau\sum_i\sinh\theta_i}\,
f(\theta_1,\theta_2,\theta_3,\theta_4)\,
\bra{\theta_4,\theta_3,\theta_2,\theta_1}\sigma^\mathrm{z}
\ket{-\xi_1,\xi_1,-\xi_2,\xi_2}.
\end{split}
\label{eq:C22}
\end{equation}
If we regularise the second matrix element according to \eqref{eq:reg1} we
first obtain (see Fig.~\ref{fig:diagrams6}) the completely disconnected term
$C_{44}^0(\tau,x_1,x_2)=C_{04}(\tau,x_2,x_1)$.  Along the same lines as in the
calculation of $C_{24}^1$ we find after some algebra that
$C_{44}^1(\tau,x_1,x_2)=C_{24}^1(\tau,x_2,x_1)$.  Furthermore, there are two
different terms with two lines connecting the two operators, first
\begin{equation}
\begin{split}
C_{44}^{2,\mathrm{a}}(\tau,x_1,x_2)=
&\,-\frac{\sigma_0^2}{2}\int_0^\infty\frac{d\xi_1d\xi_2}{(2\pi)^2}
\int_{-\infty}^\infty\frac{d\theta_1 d\theta_2}{(2\pi)^2}\,
\prod_{i=1}^2 K(\xi_i)\,\tanh\xi_i\,
\prod_{i=1}^{2}
\frac{\cosh\xi_1-\ii\sinh\theta_i}{\cosh\xi_1+\ii\sinh\theta_i}\,
\frac{\cosh\xi_2+\ii\sinh\theta_i}{\cosh\xi_2-\ii\sinh\theta_i}\\
&\hspace{25mm}\cdot
\tanh^2\frac{\theta_1-\theta_2}{2}\,
e^{2Mx_1\cosh\xi_1}\,e^{2Mx_2\cosh\xi_2}\,e^{-\ii Mr\sum_i\sinh\theta_i}\,
e^{-M\tau\sum_i\cosh\theta_i},
\end{split}
\end{equation}
which is not symmetric under $\xi_1\leftrightarrow\xi_2$, and second,
\begin{equation}
\begin{split}
C_{44}^{2,\mathrm{b}}(\tau,x_1,x_2)=
&\,-\frac{\sigma_0^2}{4}\int_{-\infty}^\infty
\frac{d\xi_1d\xi_2}{(2\pi)^2}\frac{d\theta_1 d\theta_2}{(2\pi)^2}\,
K(\xi_1)\,K(\xi_2)\,\tanh^2\frac{\xi_1-\xi_2}{2}\,
\tanh^2\frac{\theta_1-\theta_2}{2}\\
&\hspace{10mm}\cdot
\prod_{i,j=1}^{2}\frac{\cosh\theta_j+\ii \sinh\xi_i}
{\cosh\theta_j-\ii \sinh\xi_i}\,
e^{2MR\sum_i\cosh\xi_i}\,
e^{-\ii Mr\sum_i\sinh\theta_i}\,
e^{\ii M\tau\sum_i\sinh\xi_i}\,e^{-M\tau\sum_i\cosh\theta_i}.
\end{split}
\label{eq:C224}
\end{equation}
Here we can again shift $\xi_{1,2}\rightarrow\xi_{1,2}+\ii\pi/2$. The result
will, however, only yield a negligible correction to the spectral function.

\subsection{Calculation of $\bs{C_{64}}$, $\bs{C_{26}}$ and $\bs{C_{46}}$}
\begin{figure}[t]
\begin{center}
\includegraphics[scale=0.3,clip=true]{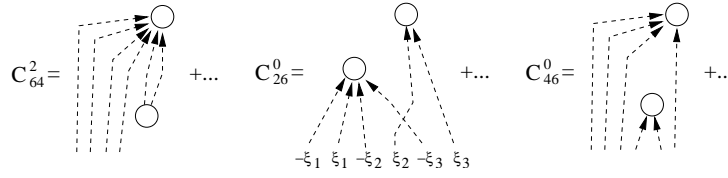}
\end{center}
\caption{Graphical representation of $C_{64}^2$, $C_{26}^0$ and
  $C_{46}^0$.}
\label{fig:diagrams7}
\end{figure}
The last terms of the series \eqref{eq:corr} we wish to calculate are the
leading contributions of $C_{64}$, $C_{26}$ and $C_{46}$ sketched in
Fig.~\ref{fig:diagrams7}. The first one is found to be
$C_{64}^2(\tau,x_1,x_2)=C_{24}^2(\tau,x_2,x_1)$.

For the evaluation of the second one we start with
\begin{equation}
\begin{split}
C_{26}(\tau,x_1,x_2)=
&\;\frac{1}{12}\int_0^\infty\frac{d\xi_1 d\xi_2 d\xi_3}{(2\pi)^3}
\int_{-\infty}^\infty\frac{d\theta_1 d\theta_2}{(2\pi)^2}\,
\prod_{i=1}^3K(\xi_i)\,e^{2Mx_2\sum_i\cosh\xi_i}\\[2mm]
&\hspace{15mm}\cdot 
e^{-Mr\sum_i\cosh\theta_i}\,e^{\ii M\tau\sum_i\sinh\theta_i}\,
f(\theta_1,\theta_2)\,
\bra{\theta_2,\theta_1}
\sigma^\mathrm{z}\ket{-\xi_1,\xi_1,-\xi_2,\xi_2,-\xi_3,\xi_3}.
\end{split}
\end{equation}
If we regularise the second matrix element according to \eqref{eq:reg1} and
keep only those terms in which the two intermediate particles are contracted
with two particles possessing different rapidities from the boundary state,
\eg the term proportional to $\bra{\theta_2,\theta_1}\xi_2,\xi_3\rangle$,  
we find
\begin{eqnarray}
C^0_{26}(\tau,x_1,x_2)&=&\ii\frac{\sigma_0^2}{2}
\int_0^\infty\frac{d\xi_1}{2\pi}\,K(\xi_1)\,\tanh\xi_1\,e^{2Mx_2\cosh\xi_1}
\nonumber\\*[2mm]
& &\hspace{10mm}\cdot
\int_{-\infty}^\infty\frac{d\xi_2 d\xi_3}{(2\pi)^2}\,
K\bigl(\xi_2+\ii\tfrac{\pi}{2}\bigr)\,K\bigl(\xi_3+\ii\tfrac{\pi}{2}\bigr)\,
\tanh^2\frac{\xi_2-\xi_3}{2}\nonumber\\*[2mm]
& &\hspace{20mm}\cdot
\prod_{i=2}^3\frac{\cosh\xi_1-\ii\sinh\xi_i}{\cosh\xi_1+\ii\sinh\xi_i}\,
e^{\ii 2MR\sum_{i=2}^3\sinh\xi_i}\,e^{-M\tau\sum_{i=2}^3\cosh\xi_i}
\label{eq:C131osc}\\*[2mm]
& &+\Theta(h_{\rm c}-h)\ 
2\ii\sigma_0^2\,\cot v\,\tan\frac{v}{2}\,e^{2MR\cos v}\,
e^{-M\tau\sin v}\nonumber\\*[2mm]
& &\hspace{5mm}\cdot
\int_0^\infty\frac{d\xi_1}{2\pi}\int_{-\infty}^\infty\frac{d\xi_2}{2\pi}\,
K(\xi_1)\,K\bigl(\xi_2+\ii\tfrac{\pi}{2}\bigr)\,
\tanh\xi_1\,e^{2Mx_2\cosh\xi_1}\,
\frac{\cosh\xi_1-\cos v}{\cosh\xi_1+\cos v}\nonumber\\*[2mm]
& &\hspace{10mm}\cdot
\tanh^2\frac{\xi_2+\ii(\pi/2-v)}{2}\,
\frac{\cosh\xi_1-\ii\sinh\xi_2}{\cosh\xi_1+\ii\sinh\xi_2}\,
e^{\ii 2MR\sinh\xi_2}\,e^{-M\tau\cosh\xi_2}.
\label{eq:C131bbs}
\end{eqnarray}
We label \eqref{eq:C131osc} by $C^{0,\mathrm{osc}}_{26}$ and
\eqref{eq:C131bbs} by $C^{0,\mathrm{bbs}}_{26}$.  In the same way one obtains
$C^0_{46}(\tau,x_1,x_2)=C^0_{26}(\tau,x_2,x_1)$.

\section{Higher-order corrections to the spectral function}
\subsection{Two- and four-particle continuum}\label{sec:appS2K}
The second-order contribution to the two-particle continuum is given by
$C^2_{24}+C^{2,\mathrm{a}}_{44}+C^2_{64}$, which after Fourier transformation
\eqref{eq:FTdef} and analytic continuation reads
\begin{equation}
\begin{split}
C^2_{2K}(\omega,R,r=0)&=
\bigl(C^2_{24}+C^{2,\mathrm{a}}_{44}+C^2_{64}\bigr)(\omega,R,r=0)\\*[2mm]
&=\frac{\sigma_0^2}{2}\int_0^\infty\frac{d\xi_1d\xi_2}{(2\pi)^2}
\int_{-\infty}^{\infty}\frac{d\theta_1 d\theta_2}{(2\pi)^2}\,
\prod_{i=1}^2 K(\xi_i)\,\tanh\xi_i\,
\frac{e^{2MR\sum_i\cosh\xi_i}}{\omega-M\sum_i\cosh\theta_i+\ii\delta}\\*[2mm]
&\hspace{20mm}\cdot
\tanh^2\frac{\theta_1-\theta_2}{2}\,
\Biggl[
\left(\frac{\cosh\xi_1-\cosh\xi_2}{\cosh\xi_1+\cosh\xi_2}\right)^2
\prod_{i,j=1}^2\frac{\cosh\xi_i+\ii\sinh\theta_j}
{\cosh\xi_i-\ii\sinh\theta_j}\\*
&\hspace{55mm}
+\prod_{i=1}^2\frac{\cosh\xi_1-\ii\sinh\theta_i}{\cosh\xi_1+\ii\sinh\theta_i}
\frac{\cosh\xi_2+\ii\sinh\theta_i}{\cosh\xi_2-\ii\sinh\theta_i}
\Biggr].
\end{split}
\end{equation}
When calculating the spectral function, the contributions proportional to the
principal value vanish as the imaginary part of the terms in the squared
brackets is antisymmetric under $\theta_i\rightarrow-\theta_i$. After some
straightforward algebra we obtain
\begin{equation}
\begin{split}
S^2_{2K}(\omega,R,r=0)
&=\sigma_0^2\int_0^\infty\frac{d\xi_1d\xi_2}{(2\pi)^2}
\int_{-\theta'}^{\theta'}\frac{d\theta}{2\pi}\,
\prod_{i=1}^2\frac{\hat{K}(\xi_i)\,\tanh\xi_i}
{(\cosh^2\xi_i+\sinh^2\theta)(\cosh^2\xi_i+\sinh^2\tilde{\theta})}\,
\frac{\tanh^2\frac{\theta-\tilde{\theta}}{2}}
{\sqrt{(\omega-M\cosh\theta)^2-M^2}}\\*[2mm]
&\hspace{-20mm}\cdot
\frac{e^{2MR\sum_i\cosh\xi_i}}{(\cosh\xi_1+\cosh\xi_2)^2}\,
\Biggl\{
\bigl(\cosh^2\xi_1+\cosh^2\xi_2\bigr)\,
\prod_{i=1}^2\Bigl[\bigl(\cosh^2\xi_i-\sinh^2\theta\bigr)
\bigl(\cosh^2\xi_i-\sinh^2\tilde{\theta}\bigr)
-4\cosh^2\xi_i\,\sinh\theta\,\sinh\tilde{\theta}\Bigr]\\*
&\hspace{20mm}
+8\prod_{i=1}^2\cosh\xi_i\,\Bigl[
\bigl(\cosh^2\xi_2-\sinh^2\theta\bigr)\sinh\tilde{\theta}+
\bigl(\cosh^2\xi_i-\sinh^2\tilde{\theta}\bigr)\sinh\theta\Bigr]\Biggr\},
\end{split}
\label{eq:S2K2}
\end{equation}
where $\tilde{\theta}$ and $\theta'$ are defined in \eqref{eq:delta1}. We
stress that the integrand is exponentially suppressed for large $M|R|$ by the
factor $e^{2MR\sum_{i=1}^2\cosh\xi_i}$.

The first order correction in the boundary reflection matrix to the
four-particle continuum is given by $C_{42}^4+C_{62}^4$. The corresponding
spectral function is
\begin{equation}
\begin{split}
S^4_{1K}(\omega,R,r=0)&=
\bigl(S_{42}^4+S_{62}^4\bigr)(\omega,R,r=0)\\[2mm]
&=\frac{\sigma_0^2}{12M}\int_0^\infty\frac{d\xi}{2\pi}
\int_{\mathcal{A}(\omega)}\frac{d\theta_1 d\theta_2 d\theta_3}{(2\pi)^3}\,
\frac{\hat{K}(\xi)\,\tanh\xi}{\cosh^2\xi+\sinh^2\tilde{\theta}_4}\,
\frac{e^{2MR\cosh\xi}}{\sinh\tilde{\theta}_4}\\
&\hspace{5mm}\cdot
\prod_{i<j}^{3}\tanh^2\frac{\theta_i-\theta_j}{2}\,
\prod_{i=1}^3\tanh^2\frac{\theta_i-\tilde{\theta}_4}{2}\,
\prod_{i=1}^3\frac{1}{\cosh^2\xi+\sinh^2\theta_i}\\
&\hspace{5mm}\cdot
\Biggl\{\bigl(\cosh^2\xi-\sinh^2\tilde{\theta}_4\bigr)\,\Biggl[
\prod_{i=1}^3\bigl(\cosh^2\xi-\sinh^2\theta_i\bigr)
-4\cosh^2\xi\;\sum_{i=1}^3\bigl(\cosh^2\xi-\sinh^2\theta_i\bigr)
\prod_{j\neq i}^3\sinh\theta_j\Biggr]\\
&\hspace{15mm}
+4\cosh^2\xi\,\sinh\tilde{\theta}_4\Biggl[
4\cosh^2\xi\,\prod_{i=1}^3\sinh\theta_i
-\sum_{i=1}^3\sinh\theta_i
\prod_{j\neq i}^3\bigl(\cosh^2\xi-\sinh^2\theta_j\bigr)\Biggr]\Biggr\},
\end{split}
\end{equation}
where $\tilde{\theta}_4$ and $\mathcal{A}(\omega)$ are defined by
\begin{equation}
\tilde{\theta}_4=\text{Arcosh}
\Bigl(\frac{\omega}{M}-\sum\nolimits_i^3\cosh\theta_i\Bigr),\;
\mathcal{A}(\omega)=\Bigl\{(\theta_1,\theta_2,\theta_3)\in\mathbb{R}^3\Big|
\sum\nolimits_i^3\cosh\theta_i\le\frac{\omega}{M}-1\Bigr\}.
\label{eq:deftildetheta4}
\end{equation}

\subsection{Oscillating terms}\label{sec:apposc}
The first correction to \eqref{eq:osc1} is given by
$C_{24}^{1,\mathrm{osc}}+C_{44}^{1,\mathrm{osc}}$. The usual steps yield for
the corresponding spectral function
\begin{equation}
\label{eq:osc3}
\begin{split}
&\bigr(S_{24}^{1,\mathrm{osc}}+S_{44}^{1,\mathrm{osc}}\bigr)(\omega,R,r=0)\\*
&\hspace{10mm}=2\sigma_0^2
\int_0^\infty\frac{d\xi}{2\pi}\int_{-\theta'}^{\theta'}\frac{d\theta}{2\pi}\,
\frac{\hat{K}(\xi)\,\tanh\xi}{\cosh^2\xi+\sinh^2\tilde{\theta}}\,
\frac{\cosh\theta-\cosh\tilde{\theta}}{\cosh\theta+\cosh\tilde{\theta}}\,
\frac{\cosh^2\xi-\sinh^2\theta}{\cosh^2\xi+\sinh^2\theta}\,
\frac{e^{2MR\cosh\xi}}{\sqrt{(\omega-M\cosh\theta)^2-M^2}}\\*[2mm]
&\hspace{20mm}\cdot
\biggl\{
\Bigl[\mathfrak{Re}K\bigl(\tilde{\theta}+\ii\tfrac{\pi}{2}\bigr)\,
\bigl(\cosh^2\xi-\sinh^2\tilde{\theta}\bigr)
+2\,\mathfrak{Im}K\bigl(\tilde{\theta}+\ii\tfrac{\pi}{2}\bigr)\,
\cosh\xi\,\sinh\tilde{\theta}\Bigr]\,
\cos\bigl(2MR\sinh\tilde{\theta}\bigr)\\*[2mm]
&\hspace{25mm}-\!
\Bigl[\mathfrak{Im}K\bigl(\tilde{\theta}+\ii\tfrac{\pi}{2}\bigr)\,
\bigl(\cosh^2\xi-\sinh^2\tilde{\theta}\bigr)
-2\,\mathfrak{Re}K\bigl(\tilde{\theta}+\ii\tfrac{\pi}{2}\bigr)\,
\cosh\xi\,\sinh\tilde{\theta}\Bigr]\,\sin\bigl(2MR\sinh\tilde{\theta}\bigr)
\biggr\},
\end{split}
\end{equation}
where $\tilde{\theta}$ and $\theta'$ are defined in \eqref{eq:delta1}.  We
stress that in comparison to \eqref{eq:osc1} there appears an extra factor
$\hat{K}(\xi)\,e^{2MR\cosh\xi}$, which shows that \eqref{eq:osc3} is indeed
the first correction in $K$ to \eqref{eq:osc1}. The first correction in $K$ to
\eqref{eq:osc2} is given by
\begin{equation}
\label{eq:osc4}
\begin{split}
&\bigr(S_{26}^{0,\mathrm{osc}}+S_{46}^{0,\mathrm{osc}}\bigr)(\omega,R,r=0)\\*
&\hspace{5mm}=-\sigma_0^2
\int_0^\infty\frac{d\xi}{2\pi}\int_{-\theta'}^{\theta'}\frac{d\theta}{2\pi}\,
\frac{\hat{K}(\xi)\,\tanh\xi}{\cosh^2\xi+\sinh^2\theta}\,
\frac{e^{2MR\cosh\xi}}{\cosh^2\xi+\sinh^2\tilde{\theta}}\,
\frac{\tanh^2\frac{\theta-\tilde{\theta}}{2}}
{\sqrt{(\omega-M\cosh\theta)^2-M^2}}\\*[2mm]
&\hspace{10mm}\cdot
\Biggl\{\biggl[
\Bigl(\mathfrak{Re}K\bigl(\theta+\ii\tfrac{\pi}{2}\bigr)\,
\mathfrak{Re}K\bigl(\tilde{\theta}+\ii\tfrac{\pi}{2}\bigr)-
\mathfrak{Im}K\bigl(\theta+\ii\tfrac{\pi}{2}\bigr)\,
\mathfrak{Im}K\bigl(\tilde{\theta}+\ii\tfrac{\pi}{2}\bigr)\Bigr)\\*
&\hspace{25mm}\cdot
\Bigl(\bigl(\cosh^2\xi-\sinh^2\theta\bigr)
\bigl(\cosh^2\xi-\sinh^2\tilde{\theta}\bigr)
-4\cosh^2\xi\,\sinh\theta\,\sinh\tilde{\theta}\Bigr)\\*
&\hspace{18mm}
-4\,\mathfrak{Re}K\bigl(\theta+\ii\tfrac{\pi}{2}\bigr)\,
\mathfrak{Im}K\bigl(\tilde{\theta}+\ii\tfrac{\pi}{2}\bigr)\,
\cosh\xi\,\Bigl(\bigl(\cosh^2\xi-\sinh^2\theta\bigr)\sinh\tilde{\theta}
+\bigl(\cosh^2\xi-\sinh^2\tilde{\theta}\bigr)\sinh\theta\Bigr)\biggr]\\*
&\hspace{85mm}\cdot\cos\bigl(2MR(\sinh\theta+\sinh\tilde{\theta})\bigr)\\*[2mm]
&\hspace{15mm}+2\,\biggl[\Bigl(
\mathfrak{Im}K\bigl(\theta+\ii\tfrac{\pi}{2}\bigr)\,
\mathfrak{Im}K\bigl(\tilde{\theta}+\ii\tfrac{\pi}{2}\bigr)-
\mathfrak{Re}K\bigl(\theta+\ii\tfrac{\pi}{2}\bigr)\,
\mathfrak{Re}K\bigl(\tilde{\theta}+\ii\tfrac{\pi}{2}\bigr)\Bigl)\,\cosh\xi\\*
&\hspace{25mm}\cdot
\Bigl(\bigl(\cosh^2\xi-\sinh^2\theta\bigr)\sinh\tilde{\theta}
+\bigl(\cosh^2\xi-\sinh^2\tilde{\theta}\bigr)\sinh\theta\Bigr)\\*
&\hspace{25mm}
-\mathfrak{Re}K\bigl(\theta+\ii\tfrac{\pi}{2}\bigr)\,
\mathfrak{Im}K\bigl(\tilde{\theta}+\ii\tfrac{\pi}{2}\bigr)
\Bigl(\bigl(\cosh^2\xi-\sinh^2\theta\bigr)
\bigl(\cosh^2\xi-\sinh^2\tilde{\theta}\bigr)
-4\cosh^2\xi\,\sinh\theta\,\sinh\tilde{\theta}\Bigr)\biggr]\\*
&\hspace{85mm}\cdot
\sin\bigl(2MR(\sinh\theta+\sinh\tilde{\theta})\bigr)
\Biggr\}.
\end{split}
\end{equation}
The oscillating contribution to local spectral function to second order in $K$
is given by
\begin{equation}
S_{2K}^{\mathrm{osc}}(\omega,R,r)=
S_{24}^{1,\mathrm{osc}}(\omega,R,r)+S_{44}^{1,\mathrm{osc}}(\omega,R,r)+
S_{26}^{0,\mathrm{osc}}(\omega,R,r)+S_{46}^{0,\mathrm{osc}}(\omega,R,r).
\end{equation}
We observe that the integrands are exponentially suppressed for large
distances from the boundary.

\subsection{Contributions from the boundary bound state}\label{sec:appbbs}
The second-order contribution of the boundary bound state to the
spectral function is given by
\begin{equation}
S_{2K}^\mathrm{bbs}(\omega,R,r)=
S_{24}^{1,\mathrm{bbs}}(\omega,R,r)+S_{44}^{1,\mathrm{bbs}}(\omega,R,r)+
S_{26}^{0,\mathrm{bbs}}(\omega,R,r)+S_{46}^{0,\mathrm{bbs}}(\omega,R,r),
\end{equation}
where 
\begin{equation}
\label{eq:bbs5}
\begin{split}
\bigl(S_{24}^{1,\mathrm{bbs}}+S_{44}^{1,\mathrm{bbs}}\bigr)(\omega,R,r=0)&=
4\sigma_0^2\,\cot v\,\tan\frac{v}{2}\,e^{2MR\cos v}\\*
&\hspace{-5mm}\cdot
\int_0^\infty\frac{d\xi}{2\pi}\,
\frac{\hat{K}(\xi)\,\tanh\xi\,e^{2MR\cosh\xi}}
{\sqrt{(\omega-M\sin v)^2-M^2}}\,
\frac{\cosh\xi-\cos v}{\cosh\xi+\cos v}\,
\frac{\cosh\hat{\xi}-\sin v}{\cosh\hat{\xi}+\sin v}\,
\frac{\cosh^2\xi-\sinh^2\hat{\xi}}{\cosh^2\xi+\sinh^2\hat{\xi}}
\end{split}
\end{equation}
and 
\begin{equation}
\label{eq:bbs6}
\begin{split}
&\bigl(S_{26}^{0,\mathrm{bbs}}+S_{46}^{0,\mathrm{bbs}}\bigr)(\omega,R,r=0)=
-\frac{4M^2\sigma_0^2}{\omega^2}\,\cot v\,\tan\frac{v}{2}\,e^{2MR\cos v}\\[2mm]
&\hspace{5mm}\cdot\int_0^\infty\frac{d\xi}{2\pi}\,
\frac{\hat{K}(\xi)\,\tanh\xi\,e^{2MR\cosh\xi}}
{\sqrt{(\omega-M\sin v)^2-M^2}}\,
\frac{\cosh\xi-\cos v}{\cosh\xi+\cos v}\,
\frac{1}{\cosh^2\xi+\sinh^2\hat{\xi}}\\[2mm]
&\hspace{15mm}\cdot
\Biggl\{\biggr[\mathfrak{Re}K\bigl(\hat{\xi}+\ii\tfrac{\pi}{2}\bigr)
\Bigl[\bigl(\cosh 2\hat{\xi}-\cos 2v-2\bigr)\,
\bigl(\cosh^2\xi-\sinh^2\hat{\xi}\bigr)/2
+4\,\cosh\xi\,\sinh^2\hat{\xi}\,\cos v\Bigr]\\*[2mm]
&\hspace{22mm}+\mathfrak{Im}K\bigl(\hat{\xi}+\ii\tfrac{\pi}{2}\bigr)
\,\sinh\hat{\xi}
\Bigl[\bigl(\cosh 2\hat{\xi}-\cos 2v-2\bigr)\,\cosh\xi
-2\,\cos v\,\bigl(\cosh^2\xi-\sinh^2\hat{\xi}\bigr)\Bigr]\biggr]\\*
&\hspace{95mm}\cdot\cos\bigl(2MR\sinh\hat{\xi}\bigr)\\[2mm]
&\hspace{18mm}-\biggr[\mathfrak{Re}K\bigl(\hat{\xi}+\ii\tfrac{\pi}{2}\bigr)
\,\sinh\hat{\xi}
\Bigl[2\,\cos v\,\bigl(\cosh^2\xi-\sinh^2\hat{\xi}\bigr)
-\bigl(\cosh 2\hat{\xi}-\cos 2v-2\bigr)\,\cosh\xi\Bigr]\\*[2mm]
&\hspace{20mm}+\mathfrak{Im}K\bigl(\hat{\xi}+\ii\tfrac{\pi}{2}\bigr)
\Bigl[\bigl(\cosh 2\hat{\xi}-\cos 2v-2\bigr)\,
\bigl(\cosh^2\xi-\sinh^2\hat{\xi}\bigr)/2
+4\,\cosh\xi\,\sinh^2\hat{\xi}\,\cos v\Bigr]\biggr]\\*
&\hspace{95mm}\cdot\sin\bigl(2MR\sinh\hat{\xi}\bigr)\Biggr\}.
\end{split}
\end{equation}
Here $\hat{\xi}$ is given in \eqref{eq:xihatbbs}.


\end{document}